\documentclass{Akhmedagaev}
\usepackage{tikz}
\usepackage{graphicx}
\usetikzlibrary{positioning}
\usepackage{xcolor}
\usepackage{epstopdf, epsfig}
\usepackage[T1]{fontenc}
\usepackage{upquote}

\title{Turbulent Rayleigh-B\'{e}nard convection in a strong vertical magnetic field}

\author{R. Akhmedagaev\aff{1},
O. Zikanov\aff{1}
  \corresp{\email{zikanov@umich.edu}},
D. Krasnov\aff{2}
 \and J. Schumacher\aff{2}}

\affiliation{\aff{1}University of Michigan - Dearborn, 4901 Evergreen Road, Dearborn, MI 48128-1491, USA
\aff{2}Technische Universit\"{a}t Ilmenau, Postfach 100565, Ilmenau, D-98694, Germany}

\begin{document}

\maketitle

\begin{abstract}
Direct numerical simulations are carried out to study flow structure and transport properties in turbulent Rayleigh-B\'{e}nard convection in a vertical cylindrical cell of aspect ratio one with an imposed axial magnetic field. Flows at the Prandtl number $0.025$ and the Rayleigh and Hartmann numbers up to $10^9$ and $1400$ are considered. The results are consistent with those of earlier experimental and numerical data. As anticipated, the heat transfer rate and kinetic energy are suppressed by strong magnetic field. At the same time, their growth with the Rayleigh number is found to be faster in flows at high Hartmann numbers. This behaviour is attributed to the newly discovered flow regime characterized by prominent quasi two-dimensional structures reminiscent of vortex sheets observed earlier in simulations of magnetohydrodynamic turbulence. Rotating wall modes similar to those in the Rayleigh-B\'{e}nard convection with rotation are found in flows near the Chandrasekhar linear stability limit. Detailed analysis of the spatial structure of the flows and its effect of global transport properties is reported.

\end{abstract}

\section{\textbf{Introduction}}

Combined turbulent thermal convection and magnetic fields significantly affect flow structure and transport of momentum and heat in electrically conducting fluids. The effect plays an important role in numerous systems found in technology and nature \citep{Ozoe05, Weiss14, Davidson16}. Prominent technological examples are the liquid metal batteries, growth of semiconductor crystals, and design of blankets and divertors in nuclear fusion reactors. The notable example of a natural system is the planetary dynamo. We consider the basic configuration of the Rayleigh-B\'{e}nard  convection (RBC) in a cylinder with thermally insulating sidewalls and an imposed vertical magnetic field. The quasistatic approximation of low magnetic Reynolds ${\Rey_m} \ll 1$ and Prandtl ${\it {P}_m} \ll 1$ numbers typical for laboratory experiments and industrial processes with liquid metal flows is utilized to describe the electromagnetic interactions. This implies that the applied magnetic field $\boldsymbol{\it{B}}$$_0$ is much stronger than the induced one $\boldsymbol b$ \citep{Davidson16}. The induced magnetic field is neglected in the expressions for the Lorentz force and Ohm's law. Furthermore, the induced field is assumed to adjust instantaneously to changes of velocity. The flow-field interaction is, essentially, approximated as one-way influence of the magnetic field on the flow. 

The effect of the vertical magnetic field on RBC has been studied experimentally \citep{Nakagawa57, Cioni00, Aurnou01, Burr01, King15, Zurner19_2}, numerically \citep{Liu18, Yan19, Akhmedagaev20} and theoretically \citep{Chandrasekhar61, Houchens02, Busse08}. The dynamics of the system is determined by three dimensionless control parameters: the Rayleigh, Hartmann and Prandtl numbers
\begin{equation} \label{eq0}
	{\it Ra} = \frac{g \alpha \Delta T H^3}{\nu \kappa}, \ \ \ {\it Ha} = B_0H \sqrt{{\frac{\sigma}{\rho\nu}}}, \ \ \ \Pran = \frac{\nu}{\kappa}, 
\end{equation} with the acceleration due to gravity $g$, the thermal expansion coefficient $\alpha$, the temperature difference between the horizontal boundaries of the fluid layer $\Delta T$, the height of the layer $H$, the kinematic viscosity $\nu$, the temperature diffusivity $\kappa$, the electrical conductivity $\sigma$, the mass density $\rho$ and the magnetic permeability of free space $\mu_0$. Systems with lateral walls add the aspect ratio $\Gamma = {D}/{H}$ as a parameter, where $D$ is the typical horizontal size, the diameter of the cylinder in our case.

The classical picture of the effect of magnetic field on RBC in an infinite horizontal layer includes suppression of turbulence \citep{Nakagawa57} and increase of the critical Rayleigh number ${\it Ra}$$_c$ of the convection onset. The latter effect is given by \citet{Chandrasekhar61} in the asymptotic form at high ${\it Ha}$ as ${\it Ra}_c \approx \pi^2{\it Ha}^2$. The dominance of quasi-two-dimensional (Q2D) cellular and columnar structures in the flows at ${\it Ra} > {\it Ra}_c$ within a layer with stress-free boundary conditions was observed by \citet{Yan19}. These regimes are uniquely identified by the slope of the Nusselt ${\it Nu}({\it Ra})$ and Reynolds ${\Rey}({\it Ra})$ numbers, the physical structure of the flows, and the relative relative sizes of terms in the governing equations. The flow regimes are identified as Q2D when the velocity and temperature fields are virtually independent of the axial direction outside of the boundary layers near the top and bottom walls.

Theoretical analysis of \citet{Busse08} predicts that the presence of the sidewalls leads to onset of convection at much lower ${\it Ra}$ than in an infinite layer. The heat transfer in that case is concentrated near the walls. Recent numerical simulations \citep{Liu18} of RBC in a rectangular cell at ${\it Ra} \sim {\it Ra}_c$ and ${\it Ra} < {\it Ra}_c$ confirmed existence of this so-called wall mode regime, in which significant flow and heat transfer are limited to narrow zones near the sidewalls (the wall modes). These modes are planar jets at moderate ${\it Ha}$, but have complex two-layer structure at high Hartmann numbers, higher than ${\it Ha}_c \equiv \sqrt{{\it Ra} }/{\pi} \approx1000$. Wall modes are also found in simulations of RBC in a cylindrical cell at $\Gamma = 4$, $\Pran = 0.025$, ${\it Ha} \le 1000$, ${\it Ra} = 10^7$ \citep{Akhmedagaev20}. The simulations results are consistent with the recent extensive measurements of RBC with vertical magnetic field in a cylinder with moderate ${\it Ra}$ and ${\it Ha}$ by \citet{Zurner19_1, Zurner19_2}. 

We also should mention a certain similarity between the behaviour found in RBC with vertical magnetic field and that in convection with rotation around vertical axis. Formation of Q2D structures, faster growth of heat transfer rate with ${\it Ra}$ at high rotation rates, and wall modes are observed in rotating systems \citep{Zhong91, Ecke92, Zhang19}.

The focus of our investigation is on the RBC with vertical magnetic field and sidewalls at ${\it Ra} \gg {\it Ra}_c$. High-resolution DNS of flows in a cylindrical cavity with $\Gamma = 1$, $\Pran = 0.025$, $10^7 \le {\it Ra} \le 10^9$ and $0 \le {\it Ha} \le 1400$ are performed. The work follows the experiments by \citet{Cioni00} carried out at the same $\Gamma$ and $\Pran$ (mercury) and at ${\it Ra} \le 3 \times 10^9$, ${\it Ha} \le 2000$ (the highest ${\it Ra}$ and ${\it Ha}$ achieved so far). The experiments have shown that the slope $\beta$ of the asymptotic power law ${\it Nu} \backsim {\it Ra}^\beta$ increases from $\beta \approx 0.26$ in non-magnetic flows to up to $\beta \approx 1$ at high ${\it Ha}$. The result is that the heat transfer rate in high-${\it Ra}$ flows in very strong magnetic fields becomes comparable to the rate in turbulent flows without the magnetic field. The physical mechanisms determining this behaviour remain uncertain.

To the best of our knowledge, this study is the first to analyze turbulent RBC at high ${\it Ha}$ and realistically low $\Pran$ in the framework of high resolution DNS. The only related simulations have been performed by \citet{Lim19} in a cubic convection cell at unrealistic $\Pran = 8$ with ${\it Ra}$ up to $10^{10}$ and ${\it Ha}$ up to $800$.

\section{\textbf{Presentation of the problem}}

\subsection{Physical model}

We consider a flow of an incompressible, viscous, electrically conducting fluid (a liquid metal) with constant physical properties contained in a cylinder with a uniform axial magnetic field. The governing equations are made dimensionless by using the cylinder's height $H$, the free-fall velocity $U=\sqrt{g \alpha \Delta T H}$, the external magnetic field strength $B_0$  and the imposed temperature difference $\Delta T=T_{bottom}-T_{top}$ as the scales of length, velocity, magnetic field and temperature, correspondingly. The Boussinesq and quasi-static approximations are used. The equations are 
\begin{equation} \label{eq1}
	\bnabla\bcdot\boldsymbol{u} = 0,
\end{equation}
\begin{equation} \label{eq2}
	\frac{\partial \boldsymbol{u}}{\partial t} + (\boldsymbol{u}\bcdot \bnabla)\boldsymbol{u} = -\bnabla p +  
	{\sqrt\frac{\Pran}{\it Ra}} (\bnabla^2 \boldsymbol{u} +{\it Ha}^2\left(\boldsymbol{j} \times \boldsymbol{e}_z\right)) 
	+ T\boldsymbol{e}_z,
\end{equation}
\begin{equation} \label{eq3}
	\frac{\partial T}{\partial t} + \boldsymbol{u}\bcdot\bnabla T = {\sqrt{\frac{1}{{\it Ra}\Pran}}} \bnabla^2 T,
\end{equation}
\begin{equation} \label{eq4}
	\boldsymbol{j} = -\bnabla \phi + (\boldsymbol{u} \times \boldsymbol{e}_z),
\end{equation}
\begin{equation} \label{eq5}
	\bnabla^2\phi = \bnabla \bcdot (\boldsymbol{u} \times \boldsymbol{e}_z),
\end{equation} where $p$, $\boldsymbol{u}$, $\phi$ and $T$ are the fields of pressure, velocity, electric potential, and deviation of temperature from a reference value. The top and bottom walls are maintained at constant temperatures ${\it T} = -0.5$ and ${\it T} = 0.5$, respectively. The lateral wall is thermally insulated, so having $\partial T/\partial n = 0$. No-slip boundary conditions for velocity are applied at the walls. All walls are perfectly electrically insulated which implies $\partial \phi/\partial n = 0$.

\subsection{Numerical method}

Governing equations (\ref{eq1}) $-$ (\ref{eq5}) are solved numerically using the finite difference scheme described earlier by \citet{Krasnov11, Krasnov12, Zhao12, Zikanov13}. Performance of the method in applications to convection with magnetic field was analyzed in comparison with other methods, in particular with the finite-volume approach  by \citet{Gelfgat18}. The spatial discretization is implemented in the cylindrical coordinates with the boundary conditions at the axis specified as discussed by \citet{Zikanov13}. The scheme is of the second order and nearly fully conservative in regards of the mass, momentum, kinetic energy, and electric charge conservation principles \citep{Krasnov11,Ni07}. The time discretization is semi-implicit and based on the Adams-Bashforth/Backward-Differentiation method of the second order. Implicit treatment is applied to the diffusive term in (\ref{eq3}) and the azimuthal derivative part of the Laplacian in (\ref{eq2}). The radial and axial parts of the viscous term in (\ref{eq2}) are treated explicitly. At every time step, three elliptic equations $-$ the projection method equation for pressure, the equation for temperature and the potential equation (\ref{eq5}) $-$ are solved using the FFT in the azimuthal direction and the cyclic reduction solver in the $r-z$ – plane. The computational grid is clustered toward the walls according to the coordinate transformation in the axial direction $z = \tanh(A_z \zeta)/\tanh(A_z)$ and in the radial direction $r = 0.9 \sin(\eta  \pi/2) + 0.1 \eta$. Here $-1 \le \zeta \le 1$, $0 \le \eta \le 1$ are the virtual uniform coordinates, in which the grid is uniform. The novel features that appear in the new version of the algorithm and its thorough verification are presented by \citet{Akhmedagaev20}.


All the results presented below are for the stage of a fully developed flow. The non-magnetic flows are taken as initial conditions for magnetoconvection flows. Global transport properties of momentum and heat transfer are quantified by the Reynolds number $\Rey = u_{rms}\sqrt{{\it Ra}/{\Pran}}$ with $u_{rms} = \langle u_r^2 + u_z^2 + u_\theta^2 \rangle_{V,t}$ and the Nusselt number ${\it Nu} = 1+ {\sqrt{{\it Ra}{\Pran}}\langle u_zT \rangle_{V,t}}$ where $\langle \cdot \rangle_{V,t}$ stands for volume and time averaging, correspondingly. The time averaging is performed over at least 100 convective time units.

\subsection{Grid sensitivity study}

The results of the grid sensitivity study are summarized in the supplementary materials. Verification of the model in comparison with experimental and numerical data is discussed in section \ref{results2}. Here we briefly discuss the grid requirements and the grids used in our simulations. In addition to internal structure of the flow, four boundary layers need to be accurately resolved \citep{Grossmann01,Davidson16}: the thermal boundary layer of thickness $\delta_T \approx 1/(2\it Nu)$, the viscous boundary layer with $\delta_v \approx 1/(4\sqrt{\Rey})$, the Shercliff layer with $\delta_{Sh} = 1/\sqrt{\it Ha}$ at the lateral wall, and the Hartmann layer with $\delta_{Ha} = 1/{\it Ha}$ at the top and bottom walls. 

We find that grids with $N = N_r \times N_z \times N_\theta = 192^3, 256^3$, and $384^3$ and the clustering parameter $A_z = 3.0$ are sufficient for non-magnetic flows at $10^7 \le {\it Ra} < 10^8$, $10^8 \le {\it Ra} < 10^9$, and ${\it Ra} = 10^9$, respectively. Further refinement of the grid does not lead to significant changes. At the same time, the grid sensitivity study and the comparison between the maximum grid step in the bulk with the estimates of the Kolmogorov length scale indicate that the non-magnetic flow at ${\it Ra} = 10^9$ is somewhat under-resolved. This is viewed as acceptable because the focus of our work is on flows at high Hartmann numbers.

Suppression of velocity gradients by the magnetic field allows us to alleviate the resolution requirements for the radial and azimuthal directions. Our study shows that $N_r$ and $N_\theta$ not higher than $192$ and $256$ grid points are needed to accurately resolve flow structures at  ${\it Ha} \ge 450$ with $10^7 \le {\it Ra} < 10^8$ and $10^8 \le {\it Ra} \le 10^9$, respectively. Considering the resolution along the axial coordinate, we need to take into account the Q2D character of the flow with weak axial gradients of velocity in the core (we must note that the gradients do not approach zero at high ${\it Ra}$ even at the strongest magnetic fields) and thin Hartmann boundary layers at the top and bottom walls. This allows us to rely on grids with the smaller number of points, but stronger near-wall clustering. The grid sensitivity study shows that, depending on the value of ${\it Ha}$, $N_z$ between $64$ and $256$ and $A_z$ between $3$ and $4$ securing not less than $6$ points within the Hartmann layer are sufficient.

As presented in detail in the supplementary material, our simulations show good grid convergence for ${\it Nu}$. The results for ${\Rey}$ are less satisfactory, especially in flows with strong magnetic field effect. The structure and character of the time evolution of the flow allow us attribute this, at least partially, to the effect of strong and slow (on the time scale of many tens of convective units; see, e.g. figure \ref{fig1}(e)) fluctuations of velocity. Unfeasibly large averaging times are needed to eliminate this factor.

 \section{\textbf{Results}} \label{results}
 
 \subsection{Spatial structure of the flow} \label{results1}

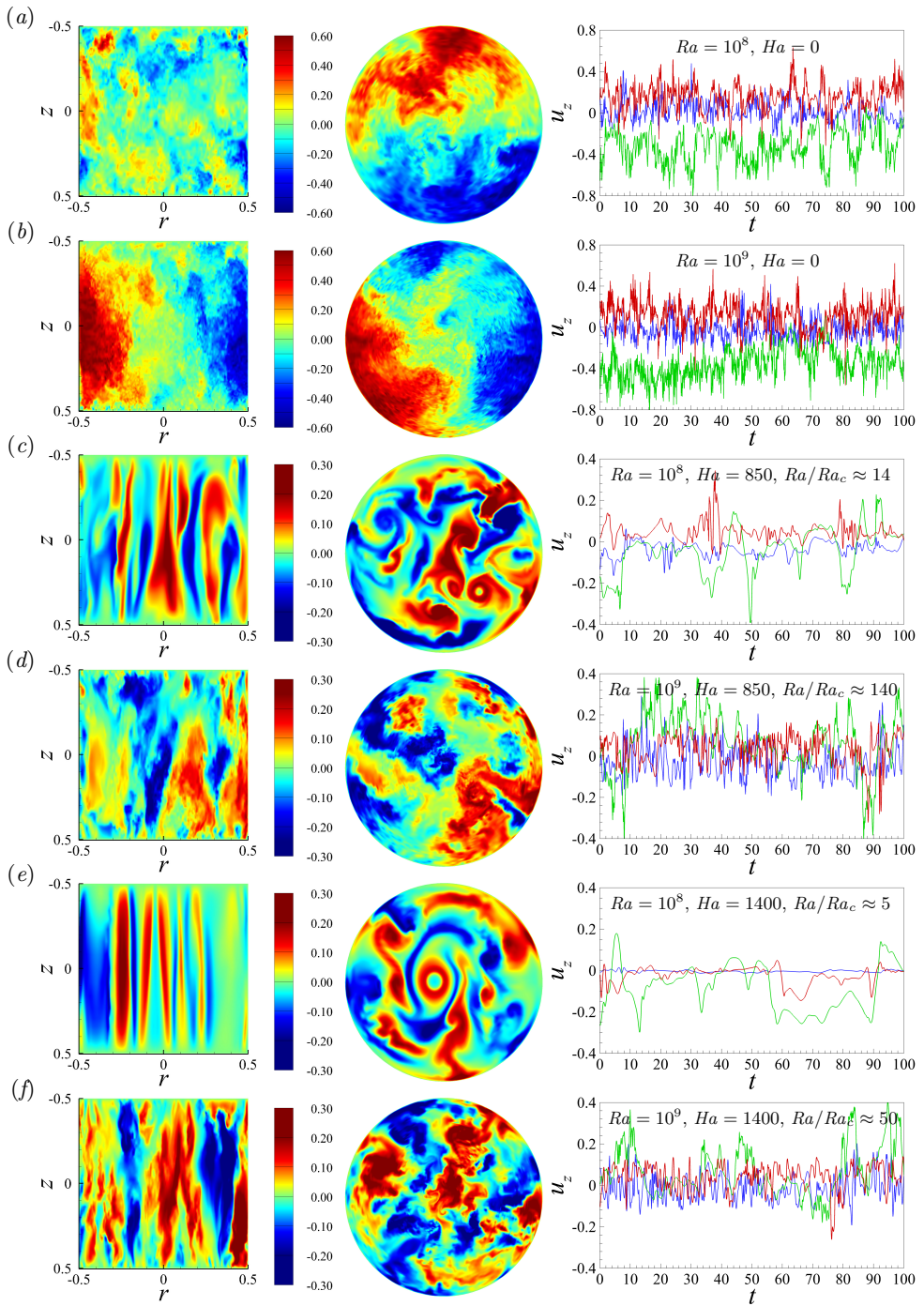
\begin{figure}
	\centering 
	
\begin{tikzpicture}

\node (img1a) {\includegraphics[scale=0.2]{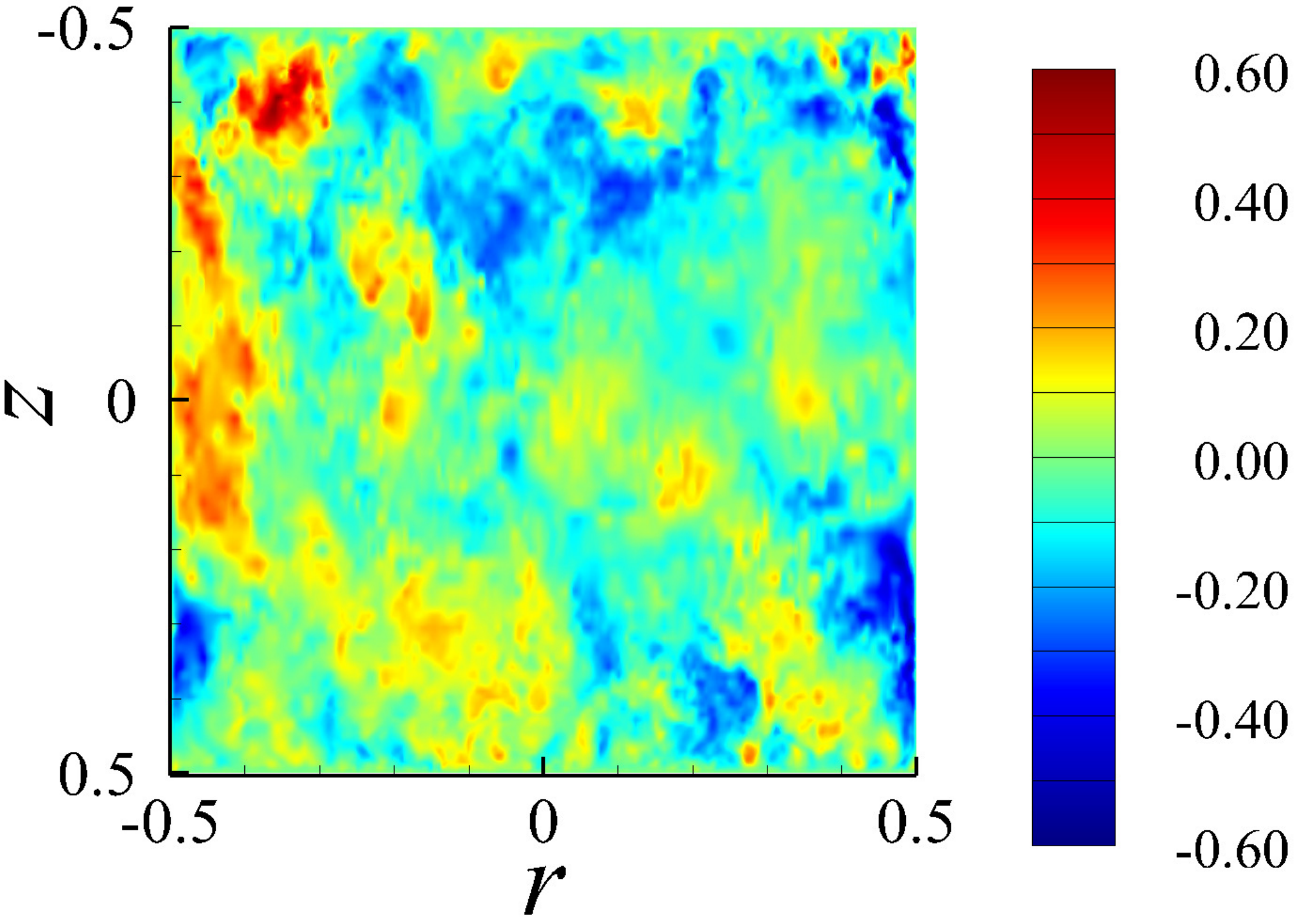}}; 
  	\node[left=of img1a, xshift=1.15cm ,yshift=1.5cm,rotate=0,font=\color{black}] {({\it a})};

\node [right=of img1a, xshift=-1.15cm, yshift=0.00cm]  (img2a)  {\includegraphics[scale=0.135]{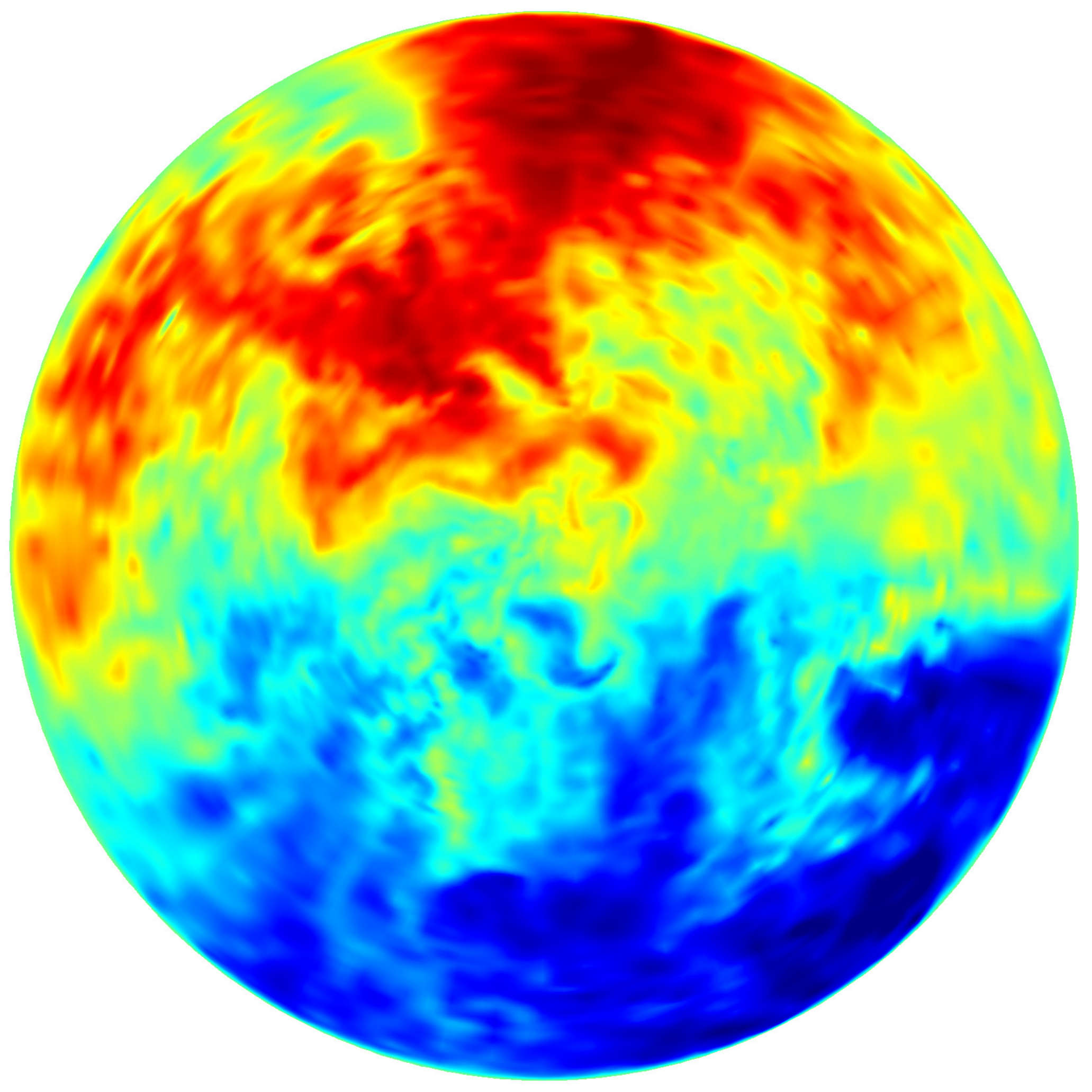}};
\node [right=of img2a, xshift=-1.15cm, yshift=-0.05cm] (img3a)  {\includegraphics[scale=0.245]{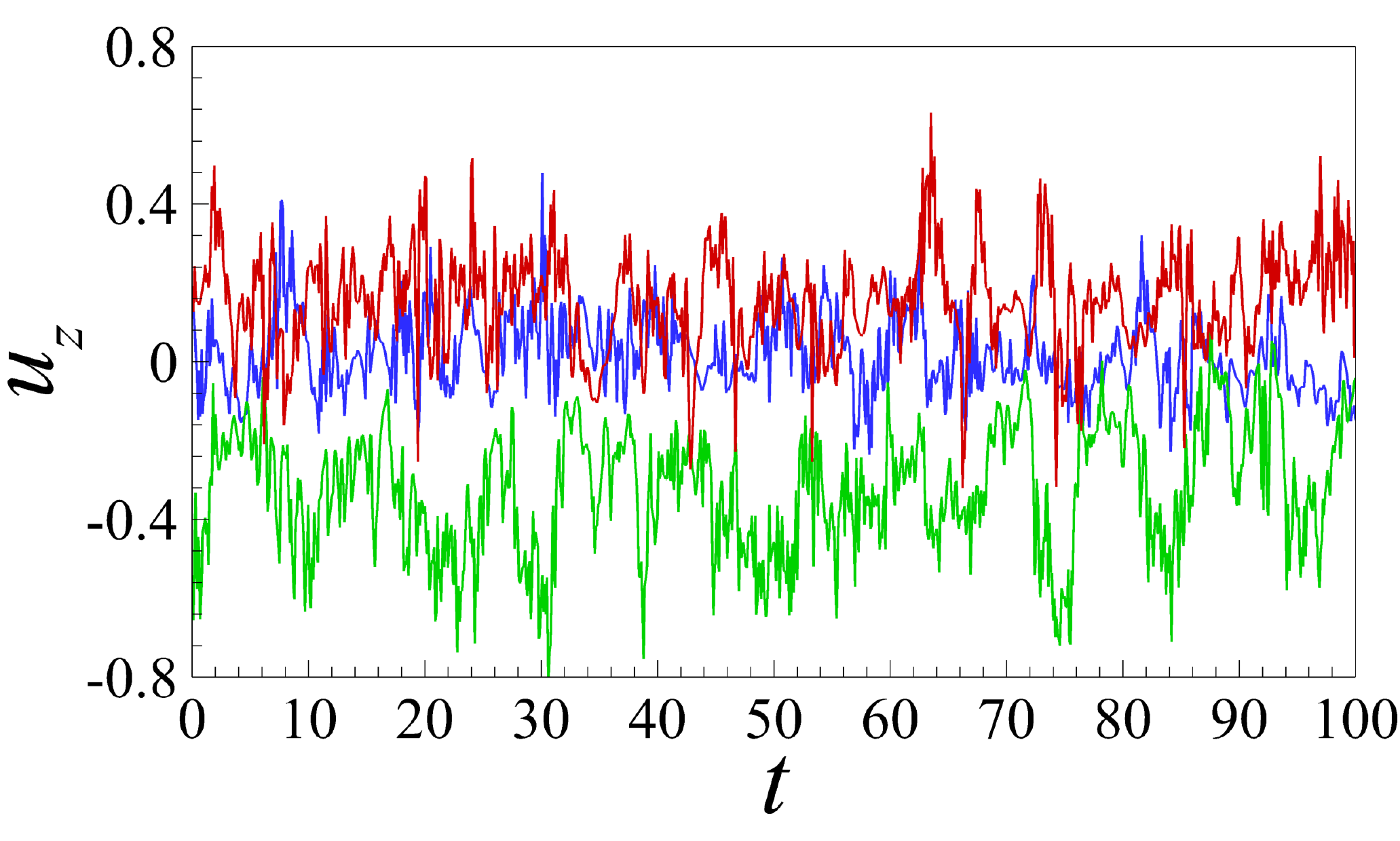}};
  	\node[left=of img3a, xshift=5.0cm ,yshift=1.15cm,rotate=0,font=\color{black},scale=0.80] {${\it Ra} = 10^8$, ${\it Ha} = 0$};

\node [below=of img1a, yshift=1.15cm] (img1b)  {\includegraphics[scale=0.200] {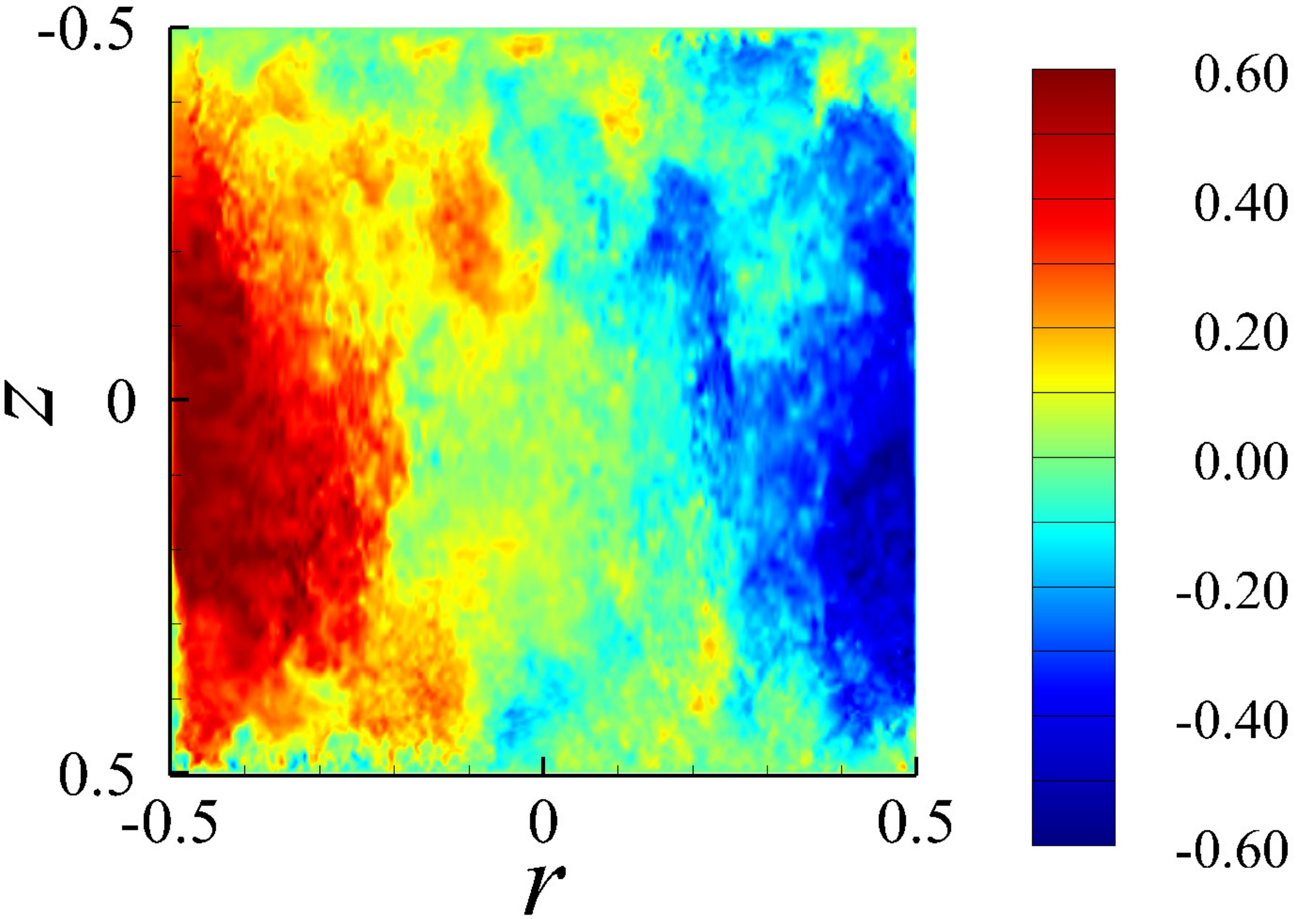}};
  	\node[left=of img1b, xshift=1.15cm ,yshift=1.5cm,rotate=0,font=\color{black}] {({\it b})}; 

\node [right=of img1b, xshift=-1.15cm, yshift=0.00cm]  (img2b)  {\includegraphics[scale=0.135] {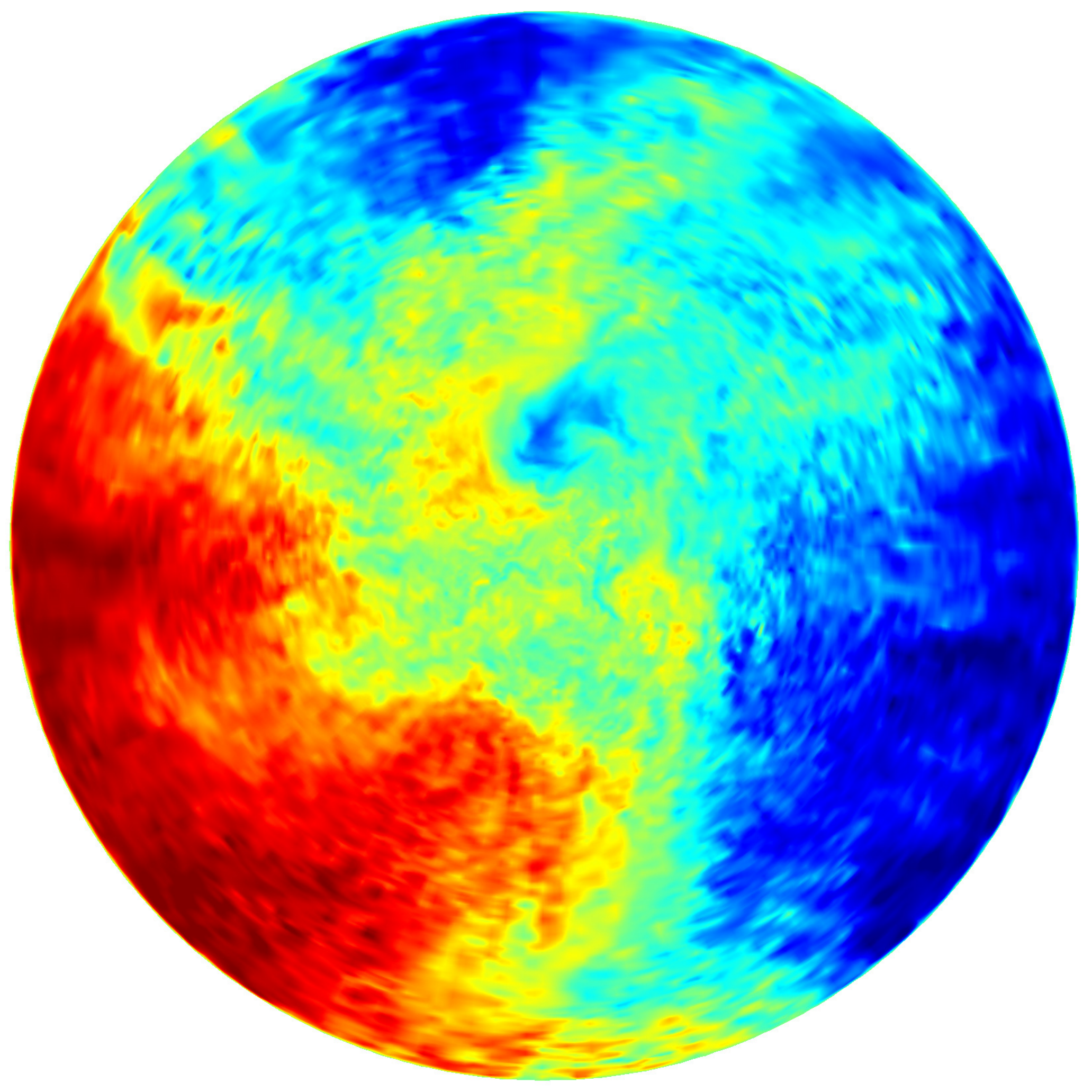}};
\node [right=of img2b, xshift=-1.15cm, yshift=-0.05cm] (img3b)  {\includegraphics[scale=0.245] {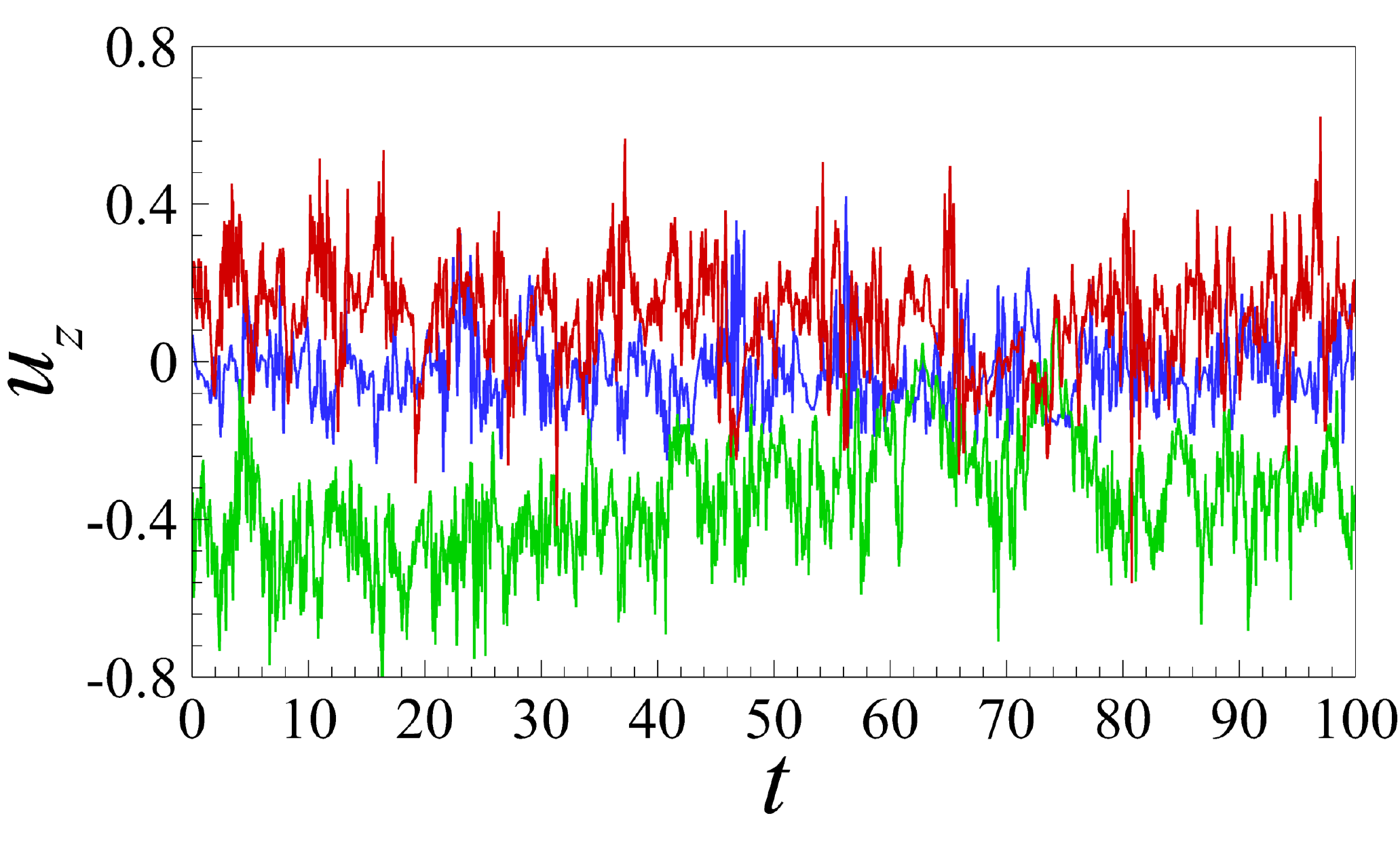}};
  	\node[left=of img3b, xshift=5.0cm ,yshift=1.15cm,rotate=0,font=\color{black},scale=0.80] {${\it Ra} = 10^9$, ${\it Ha} = 0$};

\node [below=of img1b, yshift=1.15cm] (img1c)  {\includegraphics[scale=0.200] {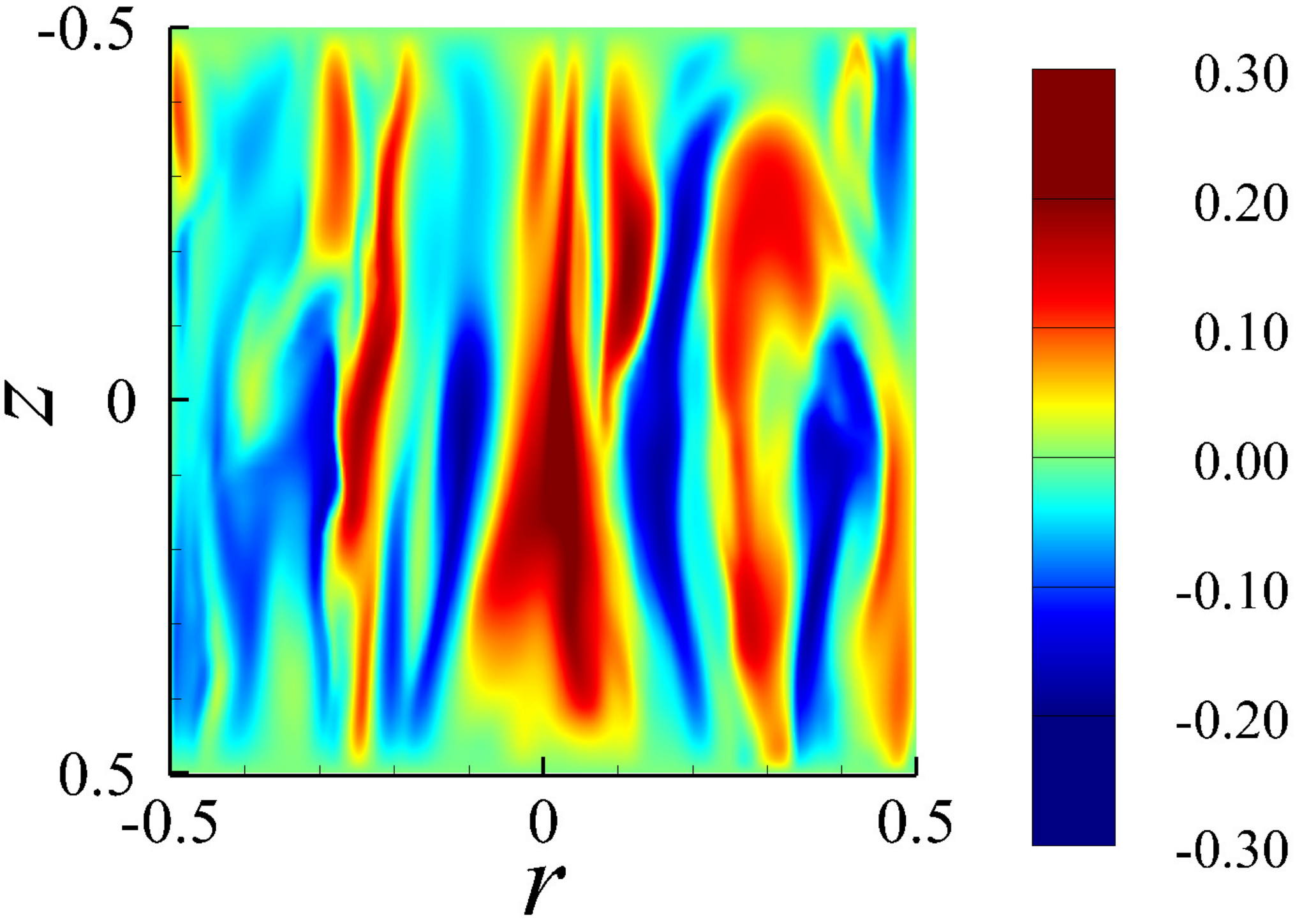}};
  	\node[left=of img1c, xshift=1.15cm ,yshift=1.5cm,rotate=0,font=\color{black}] {({\it c})}; 

\node [right=of img1c, xshift=-1.15cm, yshift=0.00cm]  (img2c)  {\includegraphics[scale=0.135] {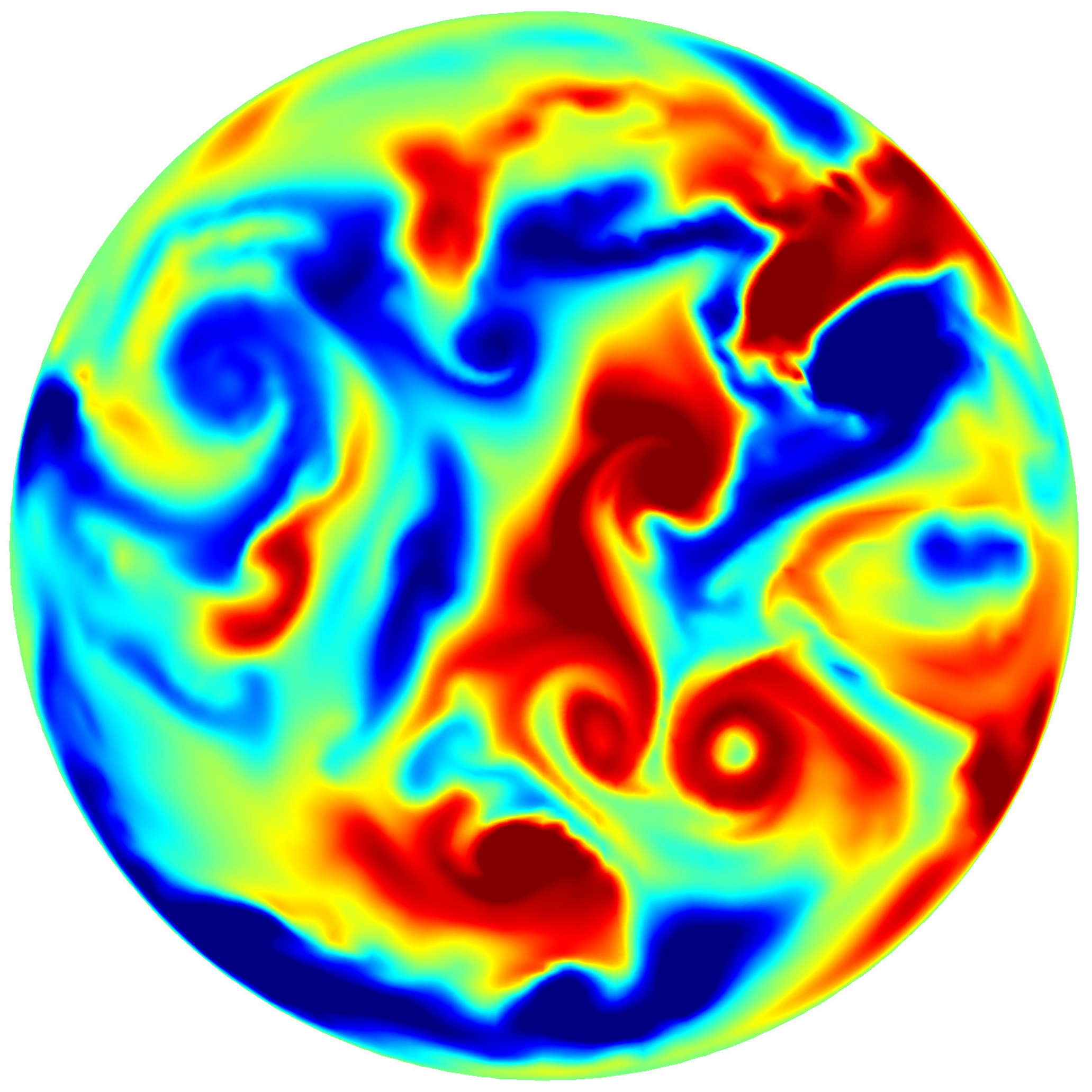}};
\node [right=of img2c, xshift=-1.15cm, yshift=-0.05cm] (img3c)  {\includegraphics[scale=0.245] {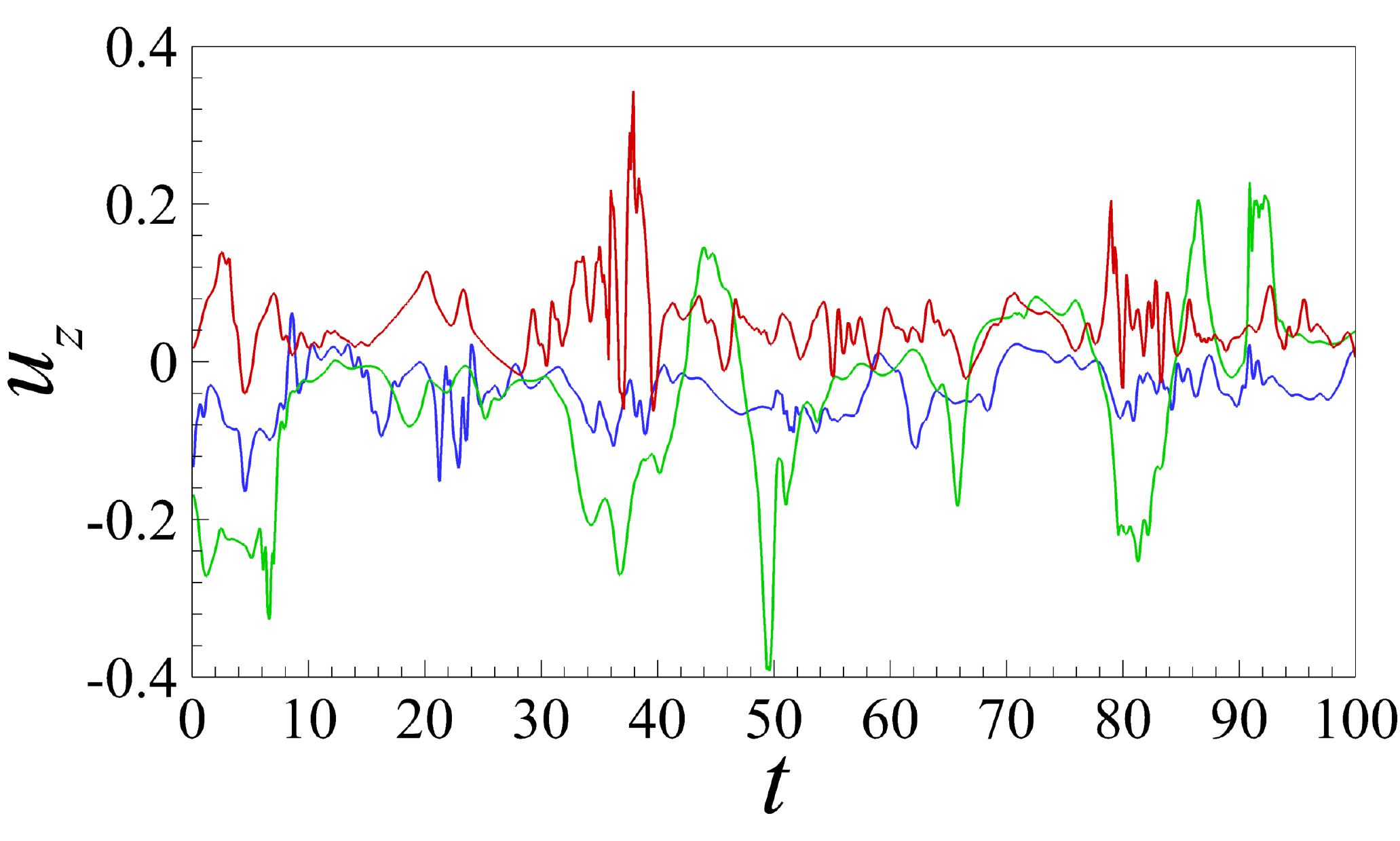}};
  	\node[left=of img3c, xshift=6.00cm ,yshift=1.15cm,rotate=0,font=\color{black},scale=0.80] {${\it Ra} = 10^8$, ${\it Ha} = 850$, ${\it Ra}/{\it Ra}_c \approx 14$};

\node [below=of img1c, yshift=1.15cm] (img1d)  {\includegraphics[scale=0.200] {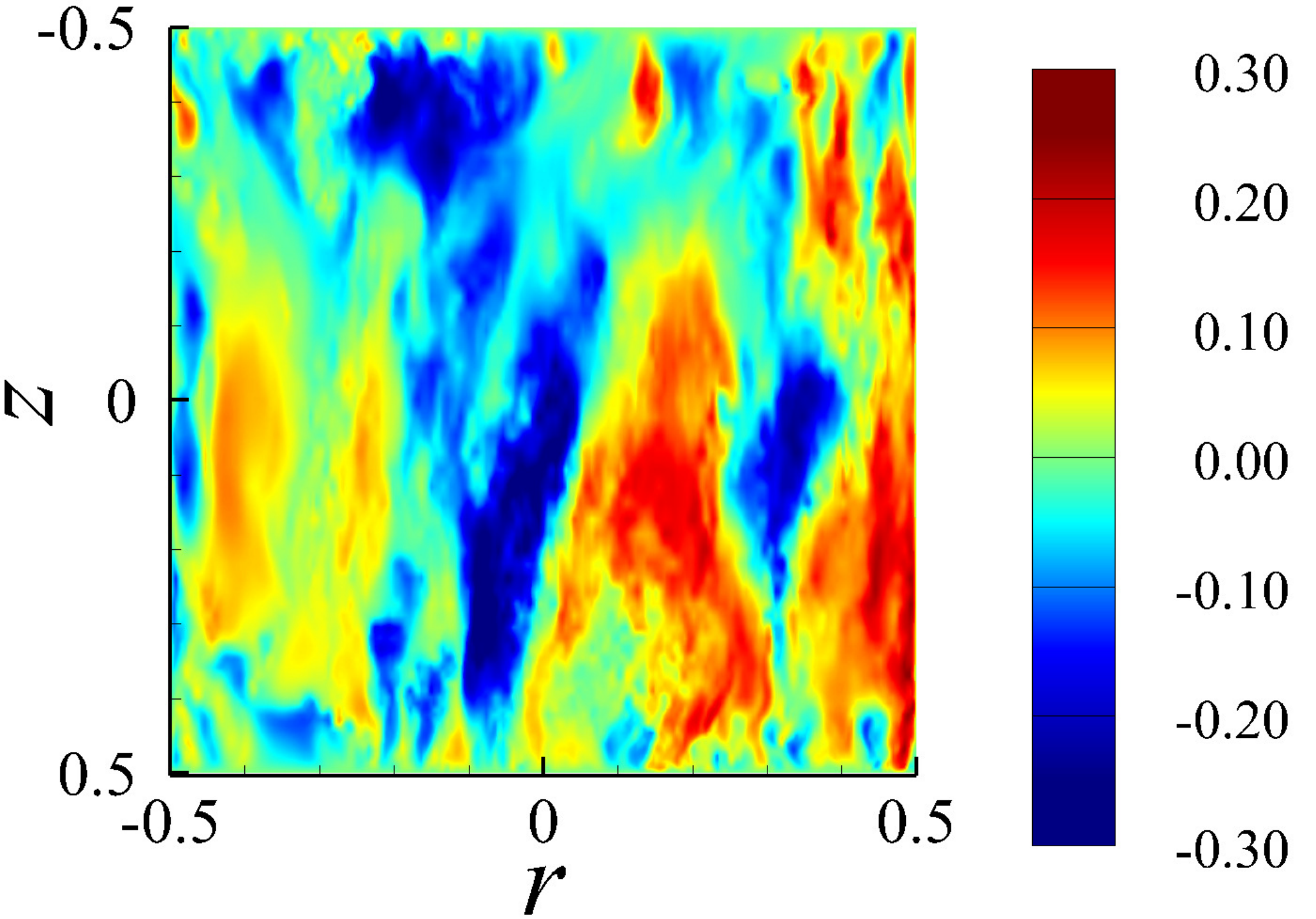}};
  	\node[left=of img1d, xshift=1.15cm ,yshift=1.5cm,rotate=0,font=\color{black}] {({\it d})}; 

\node [right=of img1d, xshift=-1.15cm, yshift=0.00cm]   (img2d)  {\includegraphics[scale=0.135] {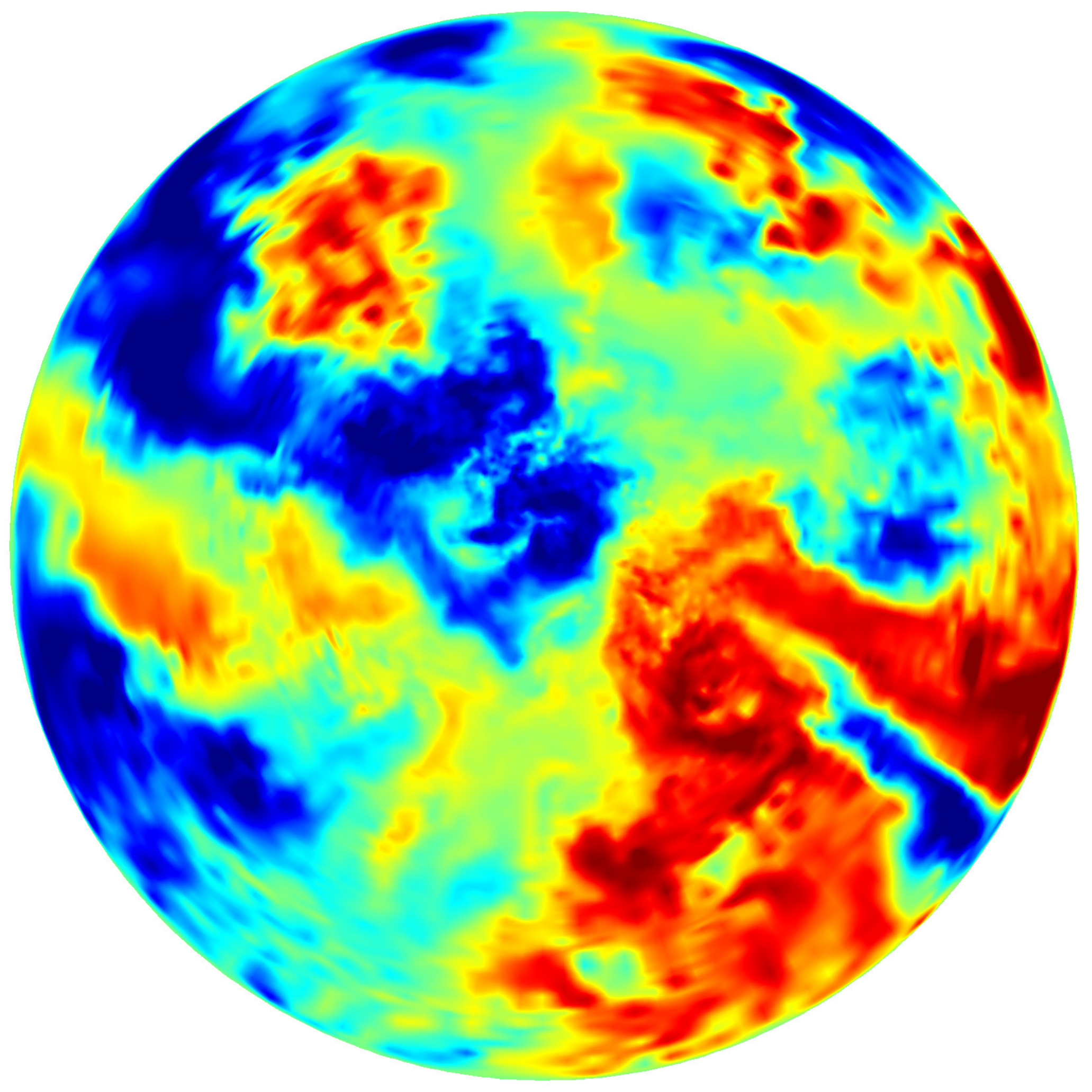}};
\node [right=of img2d, xshift=-1.15cm, yshift=-0.05cm]  (img3d)  {\includegraphics[scale=0.245]{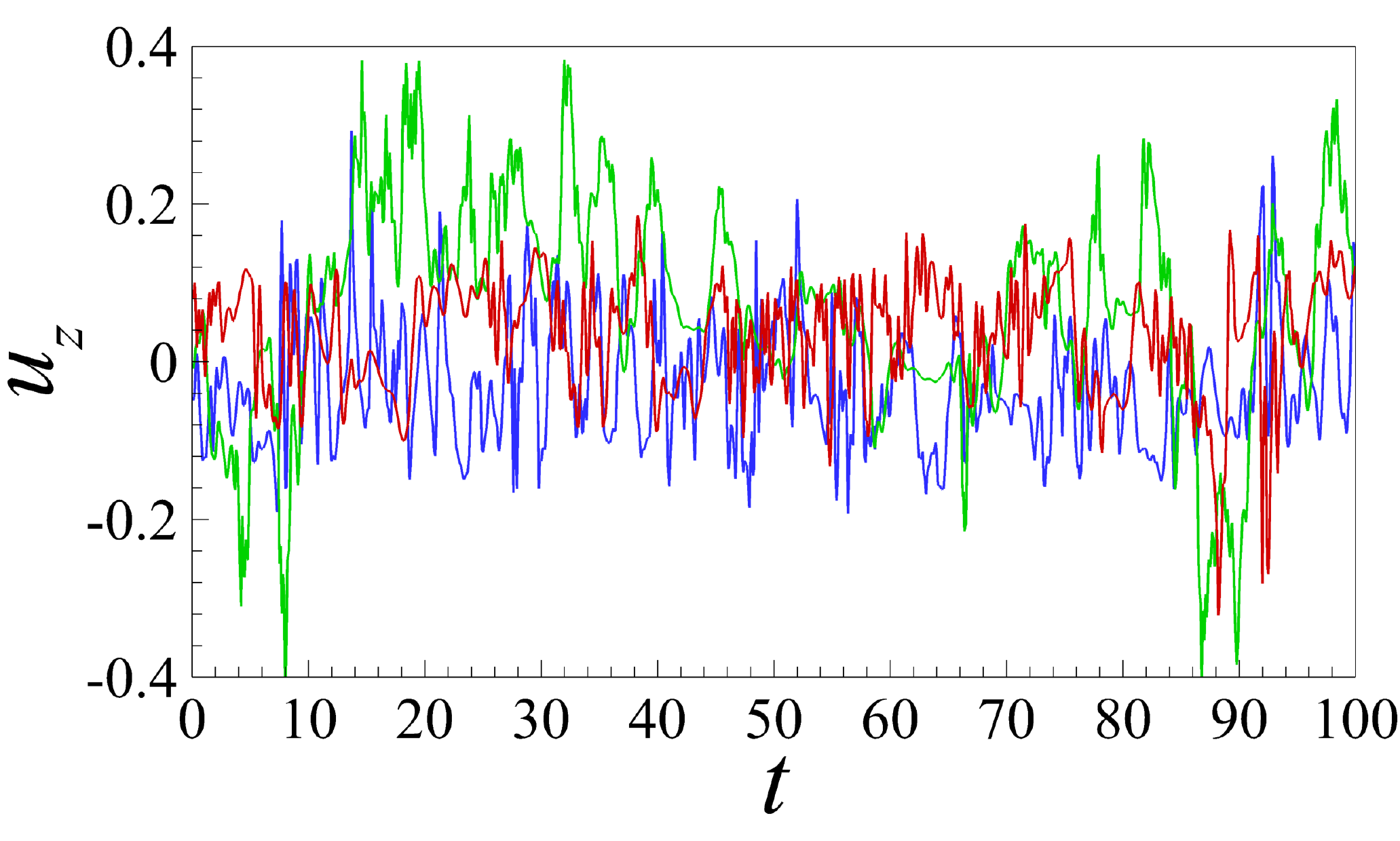}};
  	\node[left=of img3d, xshift=6.10cm ,yshift=1.15cm,rotate=0,font=\color{black},scale=0.80] {${\it Ra} = 10^9$, ${\it Ha} = 850$, ${\it Ra}/{\it Ra}_c \approx 140$};

\node [below=of img1d, yshift=1.15cm] (img1e)  {\includegraphics[scale=0.200] {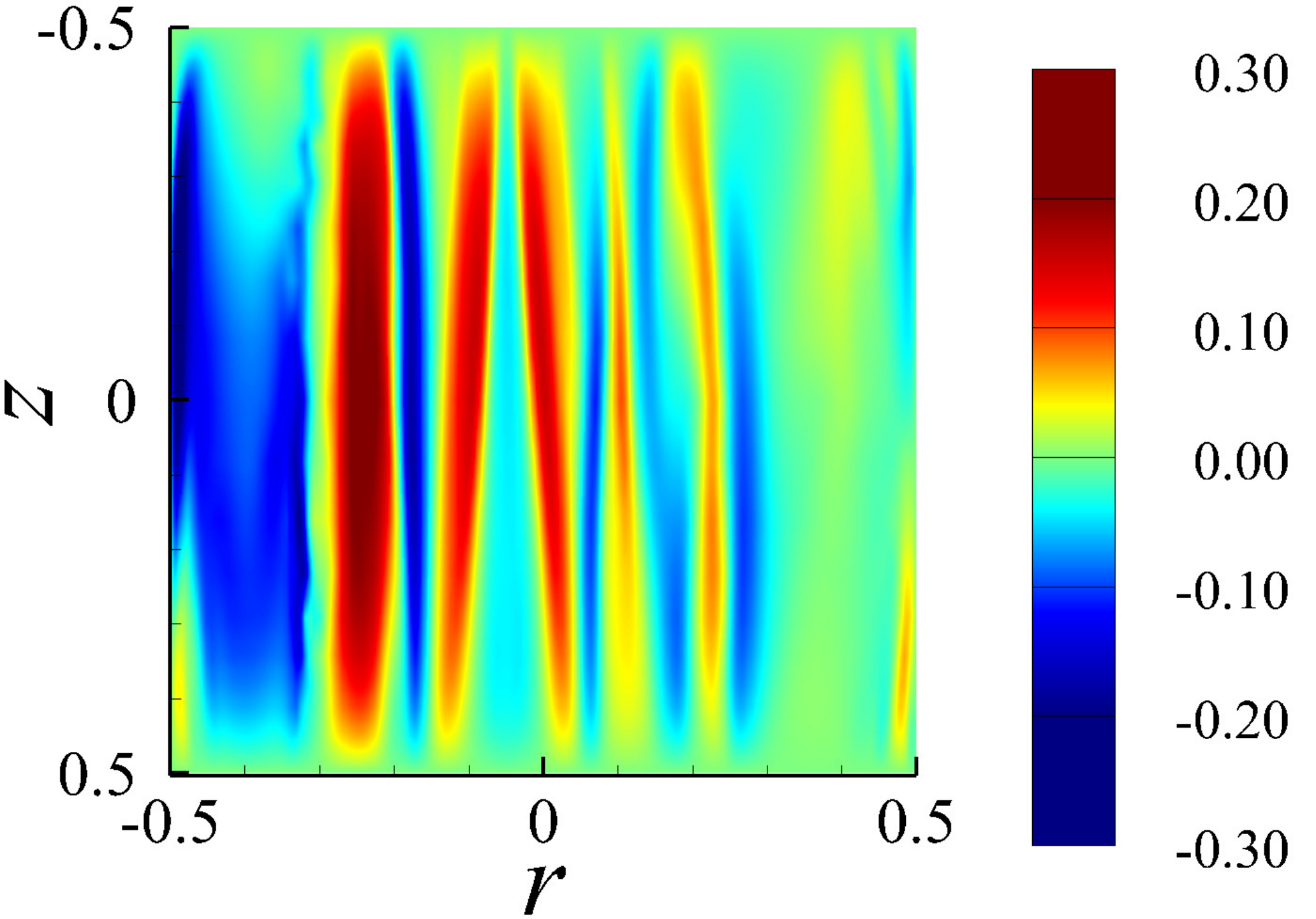}};
  	\node[left=of img1e, xshift=1.15cm ,yshift=1.5cm,rotate=0,font=\color{black}] {({\it e})}; 

\node [right=of img1e, xshift=-1.15cm, yshift=0.00cm]  (img2e)  {\includegraphics[scale=0.135] {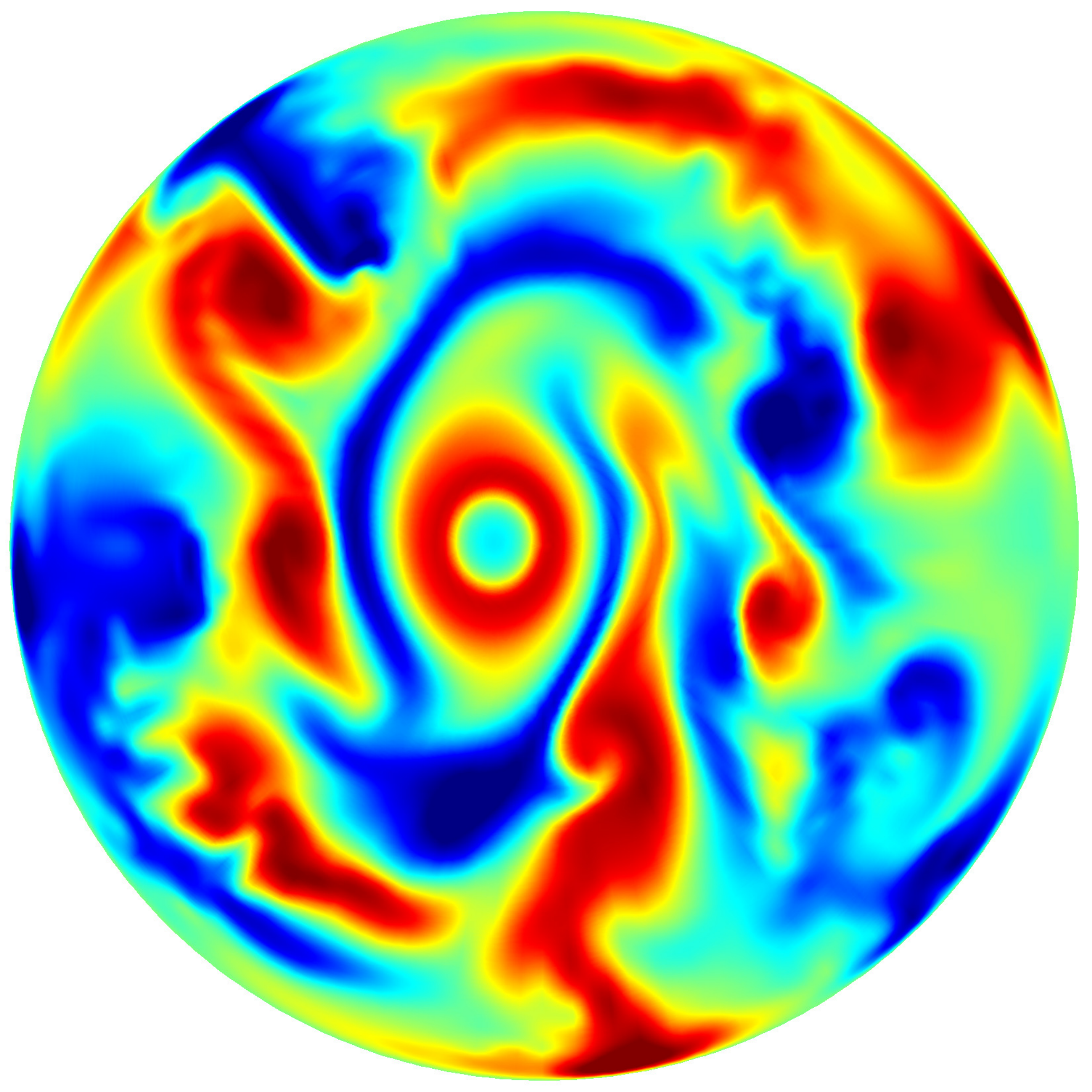}};
\node [right=of img2e, xshift=-1.15cm, yshift=-0.05cm] (img3e)  {\includegraphics[scale=0.245] {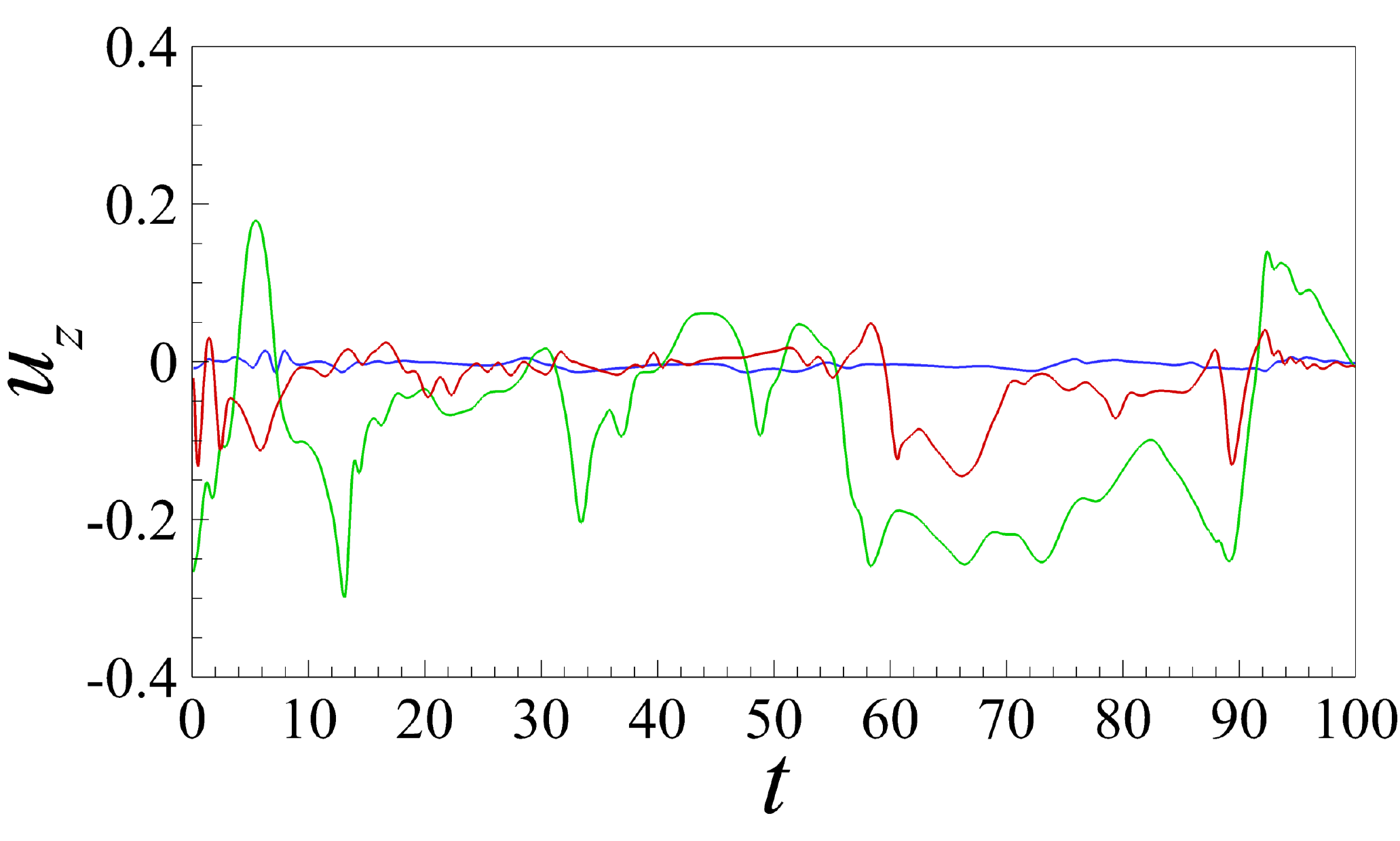}};
  	\node[left=of img3e, xshift=6.00cm ,yshift=1.15cm,rotate=0,font=\color{black},scale=0.80] {${\it Ra} = 10^8$, ${\it Ha} = 1400$, ${\it Ra}/{\it Ra}_c \approx 5$};

\node [below=of img1e, yshift=1.15cm] (img1f)  {\includegraphics[scale=0.200] {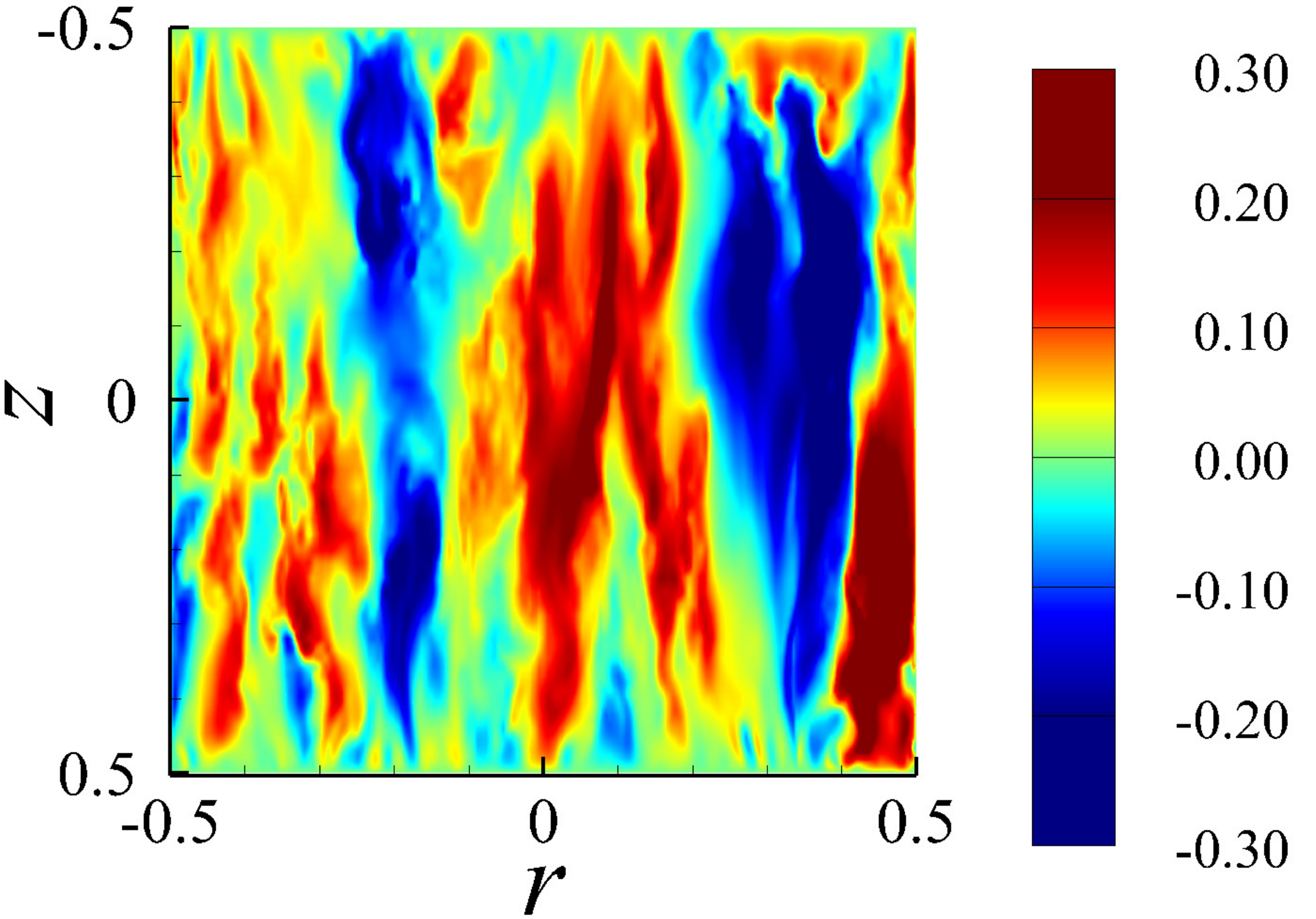}};
  	\node[left=of img1f, xshift=1.15cm ,yshift=1.5cm,rotate=0,font=\color{black}] {({\it f})}; 

\node [right=of img1f, xshift=-1.15cm, yshift=0.00cm]  (img2f)  {\includegraphics[scale=0.135] {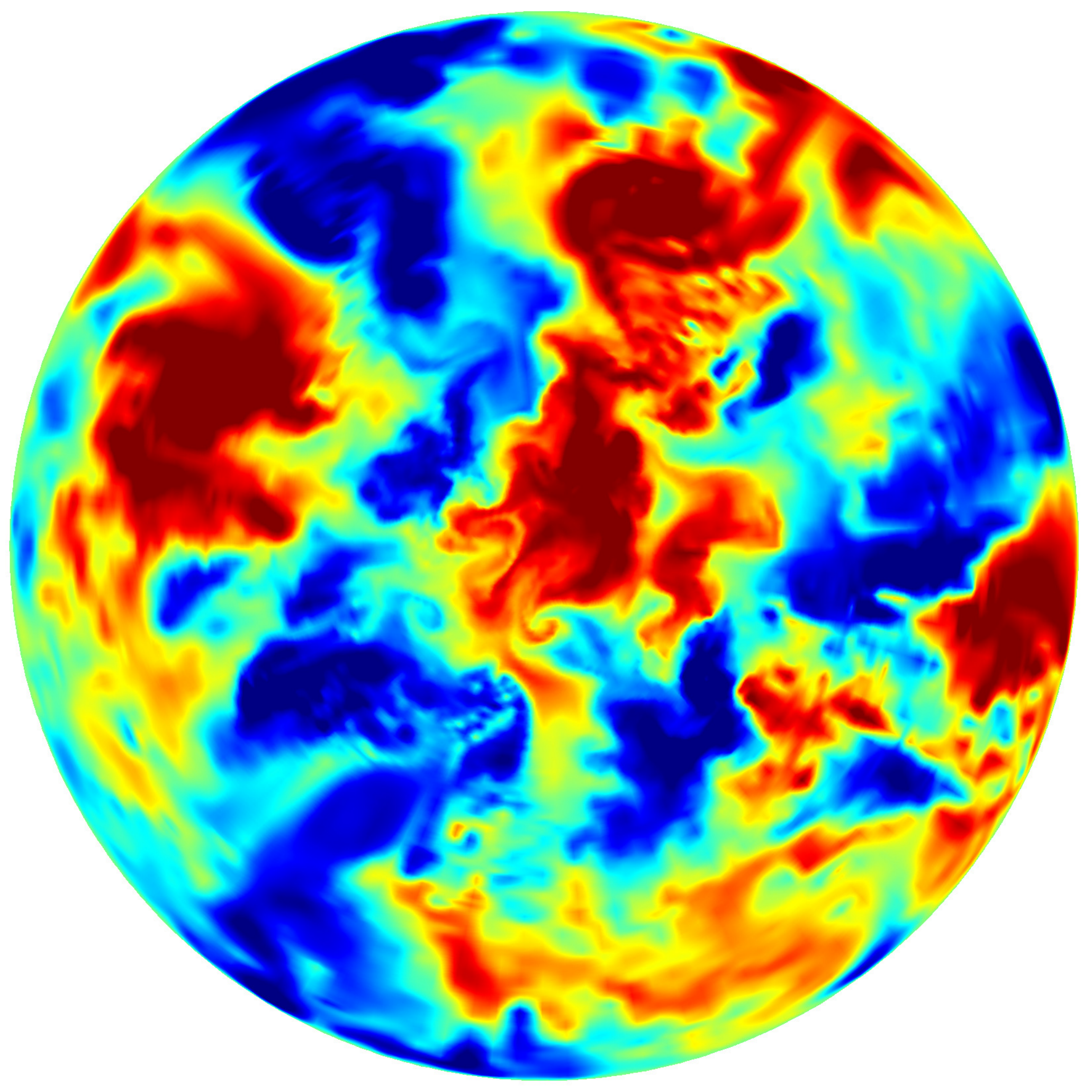}};
\node [right=of img2f, xshift=-1.15cm, yshift=-0.05cm] (img3f)  {\includegraphics[scale=0.245] {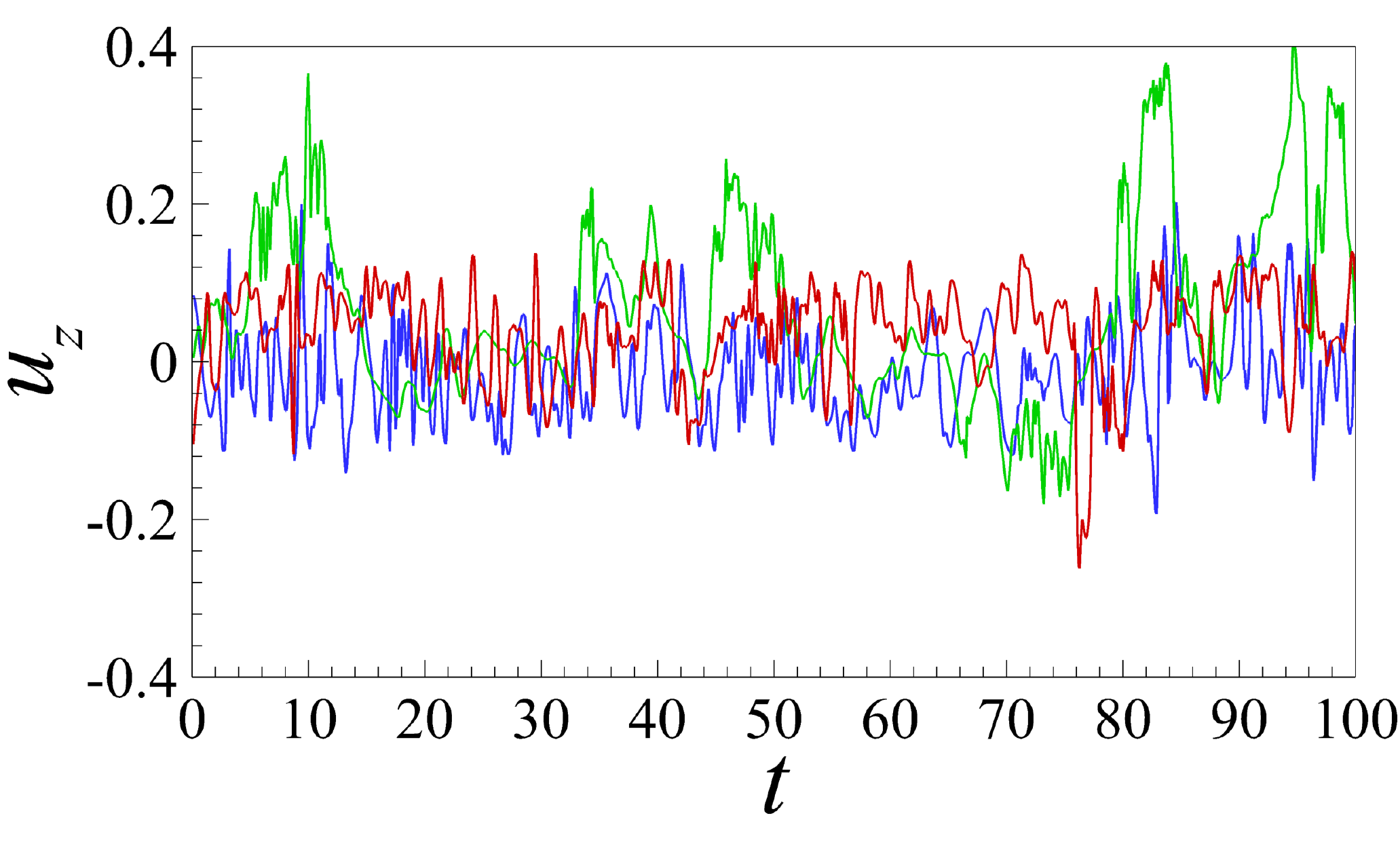}};
  	\node[left=of img3f, xshift=6.10cm ,yshift=1.15cm,rotate=0,font=\color{black},scale=0.80] {${\it Ra} = 10^9$, ${\it Ha} = 1400$, ${\it Ra}/{\it Ra}_c \approx 50$};

\end{tikzpicture}

  \caption{ Snapshots of the vertical velocity in the vertical (left column) and horizontal mid-plane (center column) cross-sections, and time signals of $u_z$ (right column) at $\theta = 0$, $z = 0.25$ (\textcolor{red}{---$\!$---}), $0$ (\textcolor{green}{---$\!$---}), $-0.25$ (\textcolor{blue}{---$\!$---}), and $r = 0.42$ ({\it e}), $0.44$ ({\it a, c, d, f}), $0.46$ ({\it b}) are shown. The parameters (${\it Ra}$, ${\it Ha}$ and the ratio between ${\it Ra}$ and ${\it Ra}_c$ of the \citet{Chandrasekhar61} stability limit) are indicated in the right column. }

  
\label{fig1}
\end{figure}

 The flow structure for several typical cases is illustrated in figure \ref{fig1}. Flows without magnetic field (see figures \ref{fig1}a and \ref{fig1}b) are turbulent. The spatial structure in the horizontal cross-section (middle column) and the correlation between the velocity signals measured along a vertical line, indicate the presence of a large-scale circulation (LSC) with upward or downward flow zones.
 
 The magnetic field drastically changes the flow. We observe suppression of small-scale velocity gradients and formation of large-scale anisotropic structures dominating the flow (see plots in the left and middle columns of figures \ref{fig1}(c)-(f)). This is associated with a reduction of amplitude and frequency of velocity fluctuations (see the plots in the right column of figures \ref{fig1}(c)-(f)) and, as we will discuss in section \ref{results2}, an increase of ${\it Nu}$ and ${\Rey}$. Similar flow transformations were detected in recent measurements of \citet{Zurner19_2} and simulations of \citet{Liu18,Yan19} for lower Rayleigh numbers.
 
 The strength of the magnetic field effect determining the degree of the flow transformation correlates with the ratio between ${\it Ra}$ and ${\it Ra}_c$. At ${\it Ra} \gg {\it Ra}_c$ (see figures \ref{fig1}d and \ref{fig1}f), the large-scale velocity structures are anisotropic (elongated in the direction of the magnetic field) but succeptible to 3D instabilities and evolve on the background of 3D small-scale velocity fluctuations. At ${\it Ra}$ larger, but not much larger than ${\it Ra}_c$ (see figures \ref{fig1}c and \ref{fig1}e), the anisotropy is much stronger and 3D fluctuations are much weaker. 
 
 In the latter case, nearly Q2D upward and downward streams occupy the flow domain. Interestingly, these flows, in paricular the flow at ${\it Ra}/{\it Ra}_c \approx 5$ shown in figure \ref{fig1}e, are principally different from the other recently identified Q2D magnetoconvection regimes: the wall modes of \citet{Liu18} and the cellular and columnar regimes found in an infinite horizontal layer by \citet{Yan19}. The former is not surprising, since the simulations of \citet{Liu18} find the wall mode regimes at ${\it Ra}/{\it Ra}_c \sim 1$ or $<1$, but not at higher ${\it Ra}/{\it Ra}_c$. For example, the wall modes were not found at ${\it Ra}/{\it Ra}_c = 4.05$.
 
 The discrepancy with the results of \citet{Yan19} obtained for an infinite horizontal layer appears more significant. Unlike the \textquotesingle{cells}\textquotesingle{} reminiscent of the linear instability modes or the \textquotesingle{columns}\textquotesingle{} found in their simulations, the dominant feature of the flow in figure \ref{fig1}e can be described as a system of ascending and descending planar jets often originating at the sidewall and extending into the bulk of the flow. We should mention that the horizontal velocity is strong in this flow ($\langle u_r^2 + u_\theta^2 \rangle_{V,t}/\langle u_z^2 \rangle_{V,t} \approx 0.44$ in comparison to $\approx 0.27$ for the case in figure \ref{fig1}f). The regime is reminiscent of Q2D extended vortex sheets often found in MHD turbulent flows with strong magnetic field effect (see, e.g., \citet{Zikanov98}). The fact that this regime rather than the cellular or columnar regimes of \citet{Yan19} is realized in our flow can be attributed  to smaller $\Pran$ ($\Pran = 1$ was used by \citet{Yan19}) or, much more likely, the presence of sidewalls.
 
\begin{figure}
	\centering 

\raisebox{11em}	
	
\begin{tikzpicture}

\node (img22a) {\includegraphics[scale=0.180]{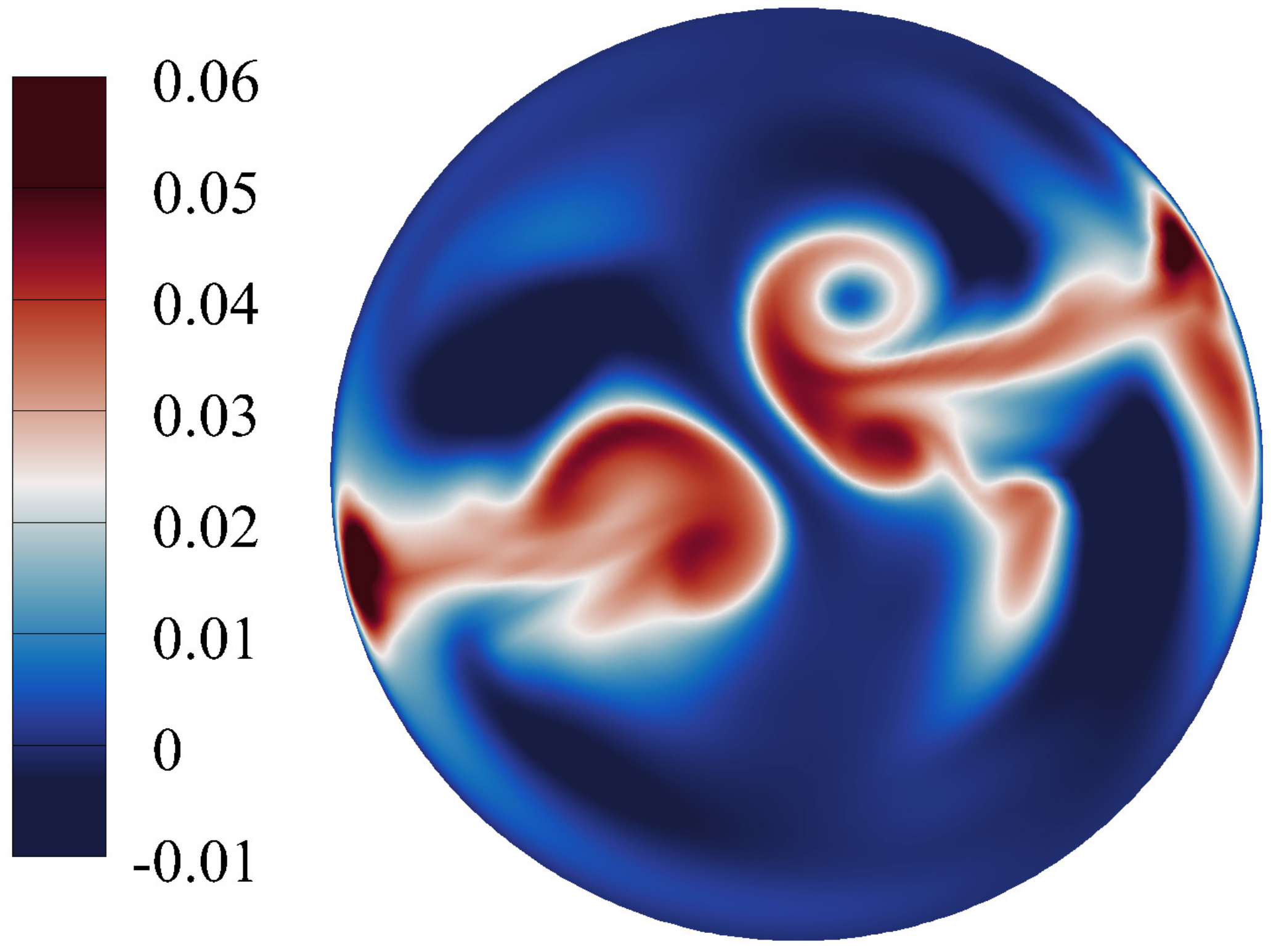}}; 
  	\node[left=of img22a, xshift=4.80cm ,yshift=1.65cm,rotate=0,font=\color{black}] {({\it a}) ${\it Ra}/{\it Ra}_c = 5.0$};
\node [right=of img22a, xshift=-1.05cm, yshift=0.00cm]  (img22b)  {\includegraphics[scale=0.135]{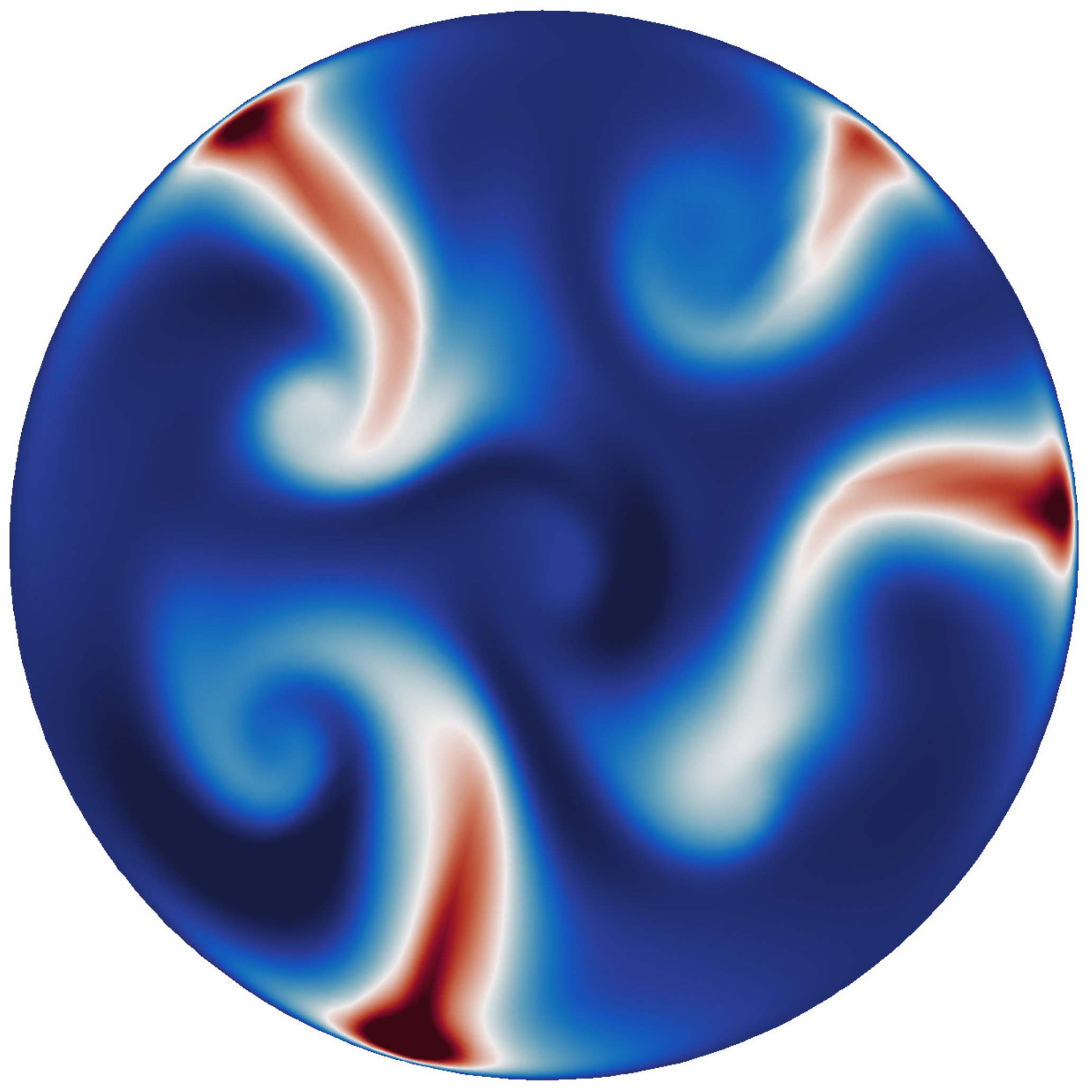}};
  	\node[left=of img22b, xshift=3.90cm ,yshift=1.65cm,rotate=0,font=\color{black}]{({\it b}) ${\it Ra}/{\it Ra}_c = 2.4$};
\node [right=of img22b, xshift=-1.05cm, yshift=-0.05cm] (img22c)  {\includegraphics[scale=0.135]{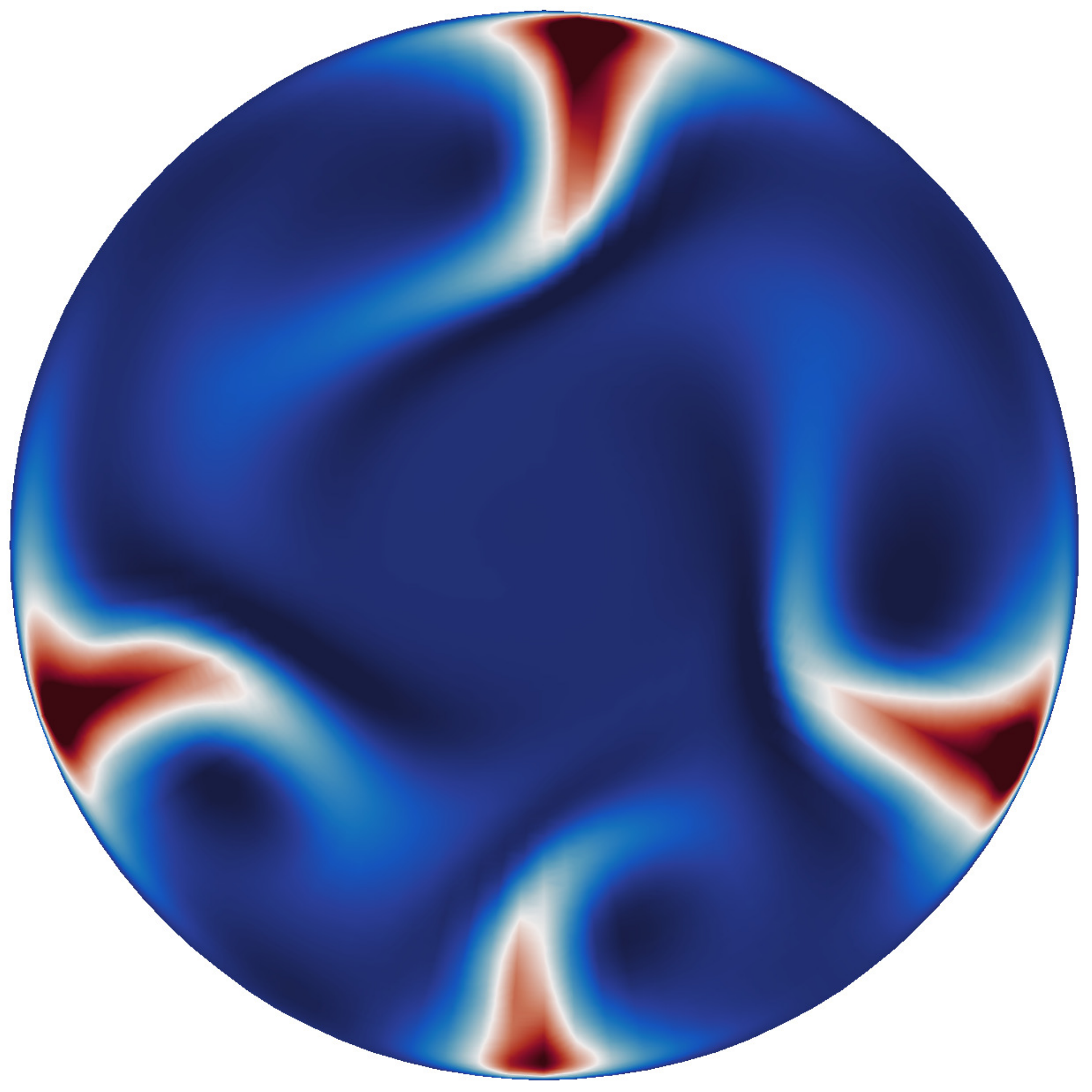}};
  	\node[left=of img22c, xshift=3.90cm ,yshift=1.70cm,rotate=0,font=\color{black}]{({\it c}) ${\it Ra}/{\it Ra}_c = 1.4$};
\node [right=of img22c, xshift=-1.05cm, yshift=-0.05cm] (img22d)  {\includegraphics[scale=0.140]{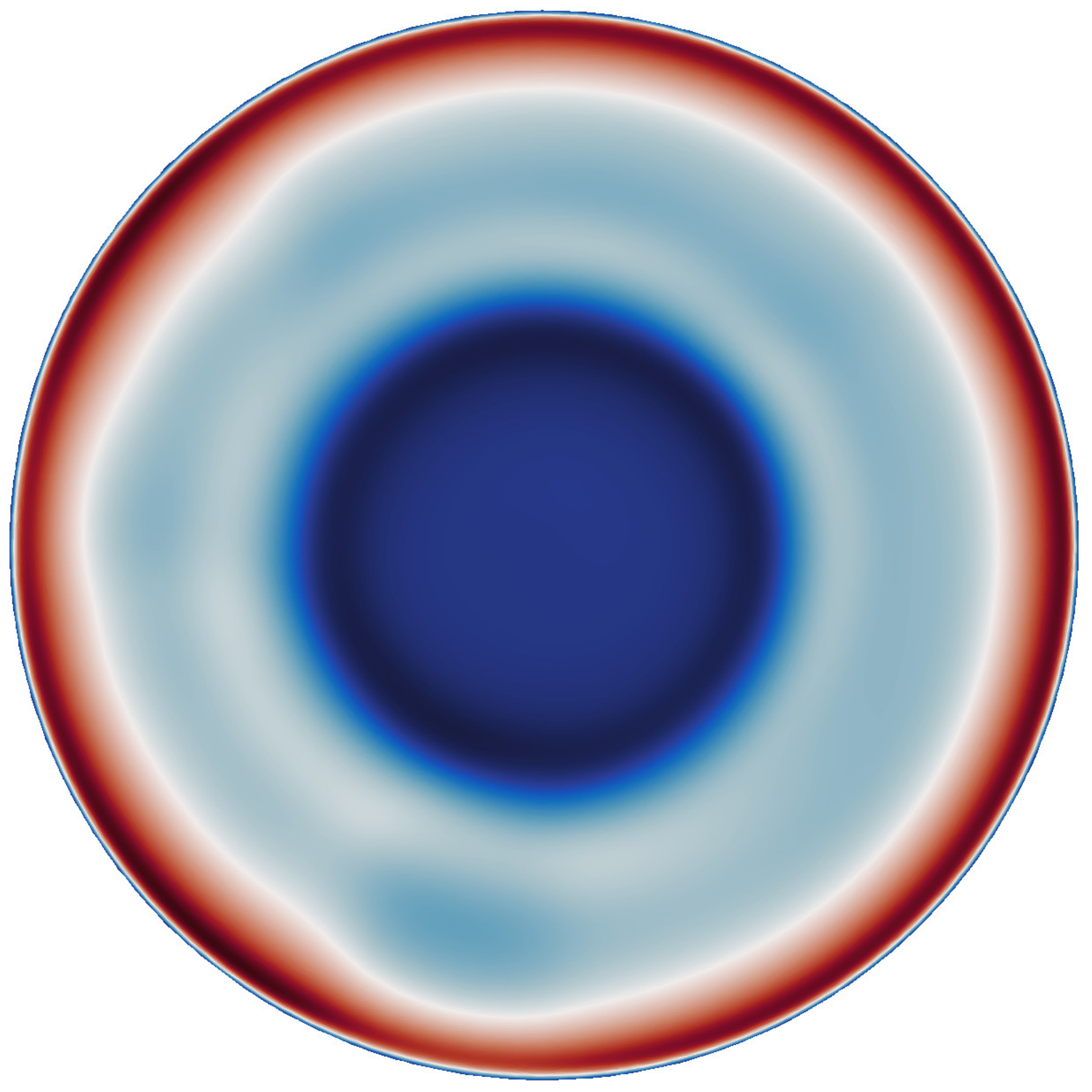}};
  	\node[left=of img22d, xshift=3.90cm ,yshift=1.72cm,rotate=0,font=\color{black}]{({\it d}) ${\it Ra}/{\it Ra}_c = 1.4$};

\end{tikzpicture}

  \caption{ Snapshots of the convective flux $u_zT$ in the horizontal mid-plane of the cylinder at ${\it Ra} = 10^7$: ${\it Ra}/{\it Ra}_c = 5.0$, ${\it Ha} = 450$ {({\it a})}, ${\it Ra}/{\it Ra}_c = 2.4$, ${\it Ha} = 650$ {({\it b})}, ${\it Ra}/{\it Ra}_c = 1.4$, ${\it Ha} = 850$ {({\it c})}. Time averaged convective flux at ${\it Ra} = 10^7$ and ${\it Ha} = 850$ is shown in {({\it d})}.}
  
\label{fig2}
\end{figure}

 The wall mode regime is found in our simulations at  ${\it Ra}/{\it Ra}_c \rightarrow 1$. As an illustration, figure \ref{fig2} shows the instantaneous distributions of convective flux $u_zT$ in flows with ${\it Ra} = 10^7$ (also used by \citet{Liu18}) and several values of ${\it Ha}$. The structure of the wall modes is qualitatively similar to the structure observed by \citet{Liu18} (see figure \ref{fig2}c). Significant flow and heat transfer are limited to tongue-like zones attached to the sidewall. There are noticeable differences, though. The entire wall-mode structure rotates clockwise or anticlockwise in our system. The effect of rotation is illustrated by the almost axisymmetric distribution of the time-averaged heat flux shown in figure \ref{fig2}d. The travelling waves in the azimuthal direction were not observed in the linear stability analysis \citep{Houchens02}. Another difference is the transformation of wall modes at increasing ${\it Ra}/{\it Ra}_c$. As in the simulations of \citet{Liu18}, the wall modes extend further from the wall. The new feature that can be plausibly attributed to rotation is that the wall-mode zones curve (see figure \ref{fig2}b) and form Q2D vortices (see figure \ref{fig2}a). We leave a detailed analysis of properties and physical mechanisms of wall modes to future studies and only mention here that observed behaviour is qualitatively similar to that in RBC with rotation \citep{Zhong91, Ecke92, Zhang19}.


 \subsection{Global transport properties} \label{results2}
 
\begin{figure}
	\centering 
	
\begin{tikzpicture}

\node (img2_1) {\includegraphics[scale=0.250]{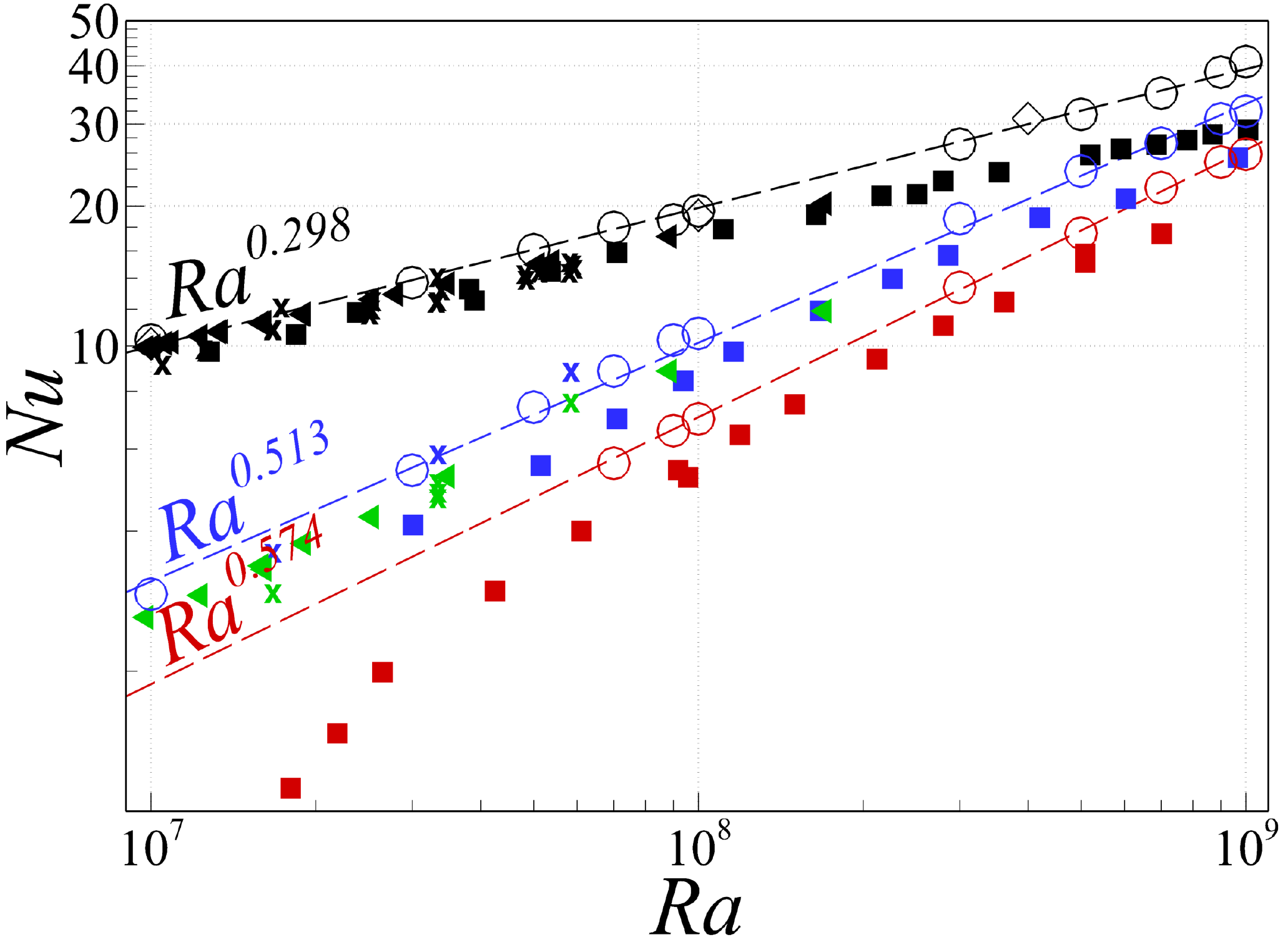}};
  	\node[left=of img2_1, xshift=2.30cm ,yshift=1.55cm,rotate=0,font=\color{black}] {({\it a})}; 
\node [right=of img2_1, xshift=-1.20cm, yshift=0.25cm] (img2_)   {\includegraphics[scale=0.125] {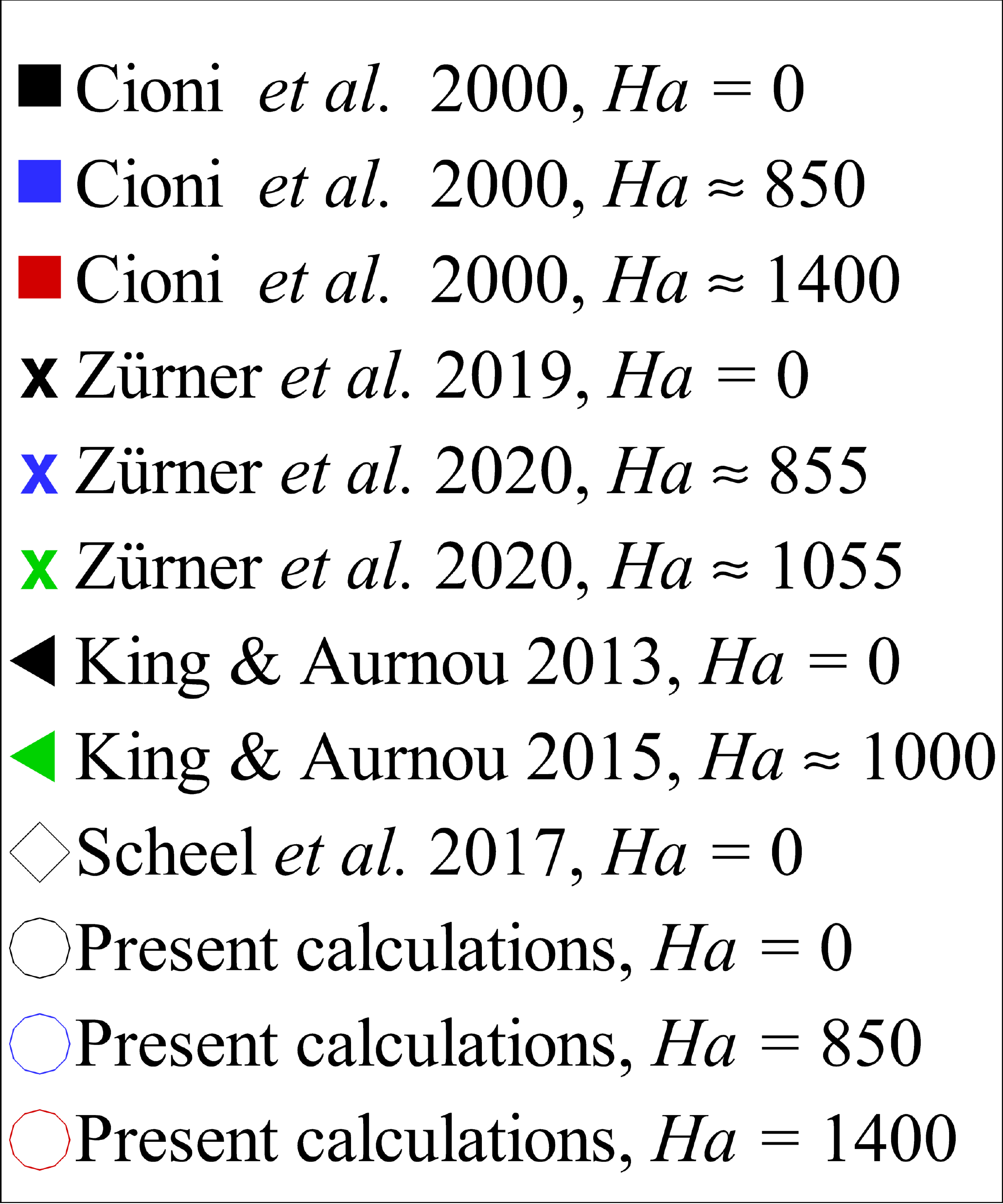}};
\node [right=of img2_1, xshift=1.45cm, yshift=0.00cm]  (img2_2)  {\includegraphics[scale=0.250]{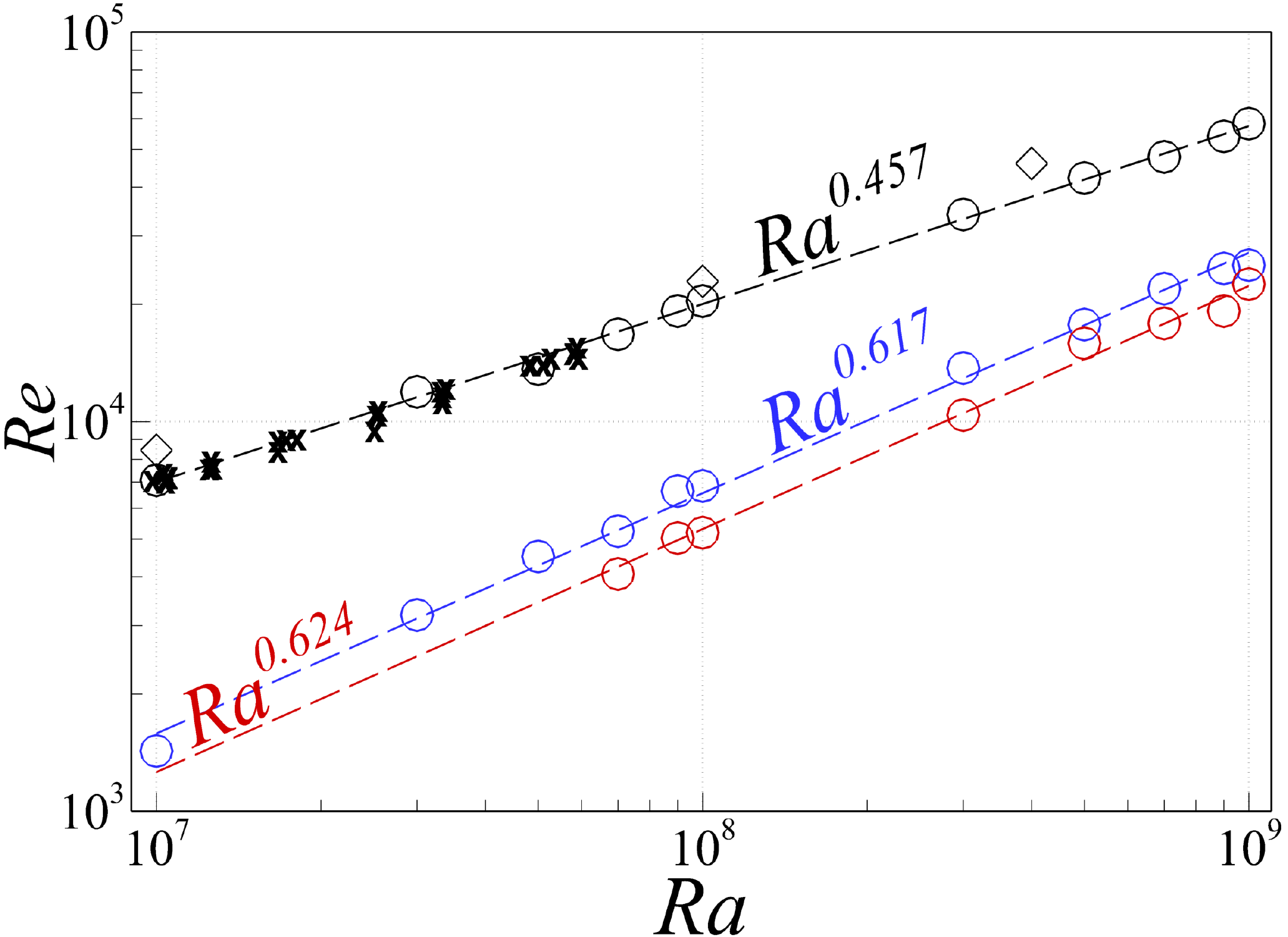}};
  	\node[left=of img2_2, xshift=2.30cm ,yshift=1.55cm,rotate=0,font=\color{black}] {({\it b})}; 

\end{tikzpicture}

  \caption{ Nusselt number ${\it Nu}$ vs ${\it Ra}$ ({\it a}) and Reynolds number ${\Rey}$ vs ${\it Ra}$ ({\it b}) with and without magnetic field. Experimental and numerical data for $\Gamma = 1$ are shown for comparison. Slope lines $ \backsim {\it Ra}^{\beta}$ are plotted for guidance.}
  
\label{fig3}  
\end{figure}

 All computed values of ${\it Nu}$ and $\Rey$ are listed in the supplementary materials. They are also summarized and compared with the available experimental and numerical data in figure \ref{fig3}. The qualitative agreement between the numerical and experimental data is good.  In particular, the increase of the slope of ${\it Nu}({\it Ra})$ and ${\Rey}({\it Ra})$ curves in flows with the magnetic field, which we discuss in detail below, is consistent between the simulations and the experiments. At the same time, the quantitative agreement is less convincing, with the computed values being consistently higher than in the experiments, especially in the experiment of \citet{Cioni00}. This situation is typical for thermal convection in low-$\Pran$ fluids and can be attributed to several known difficulties of the experimental procedure: (1) the ideal boundary conditions of constant temperature at the top and bottom plates cannot be accurately maintained; (2) temperatures at the top and bottom boundaries are not measured directly, which usually leads to a slight overestimation of the Rayleigh number; (3) complete avoidance of heat loses through the sidewall is impossible; (4) physical properties within a system are not constant, which results in inaccurate estimates of the Prandtl, Hartmann and Rayleigh numbers.
 

 The estimates of the scaling behaviour of ${\it Nu}$ and $\Rey$ based on the data of our simulations are presented in table \ref{table1}. Some of the respective lines are also shown in figure \ref{fig3}. The common regression method is applied to the computed time averaged values to determine the exponents ${\beta}_{\it Nu}$ and ${\beta}_{\it Re}$ and their standard errors, and the constants ${\alpha}_{\it Nu}$ and ${\alpha}_{\it Re}$ in the approximations ${\it Nu} \approx {\alpha}_{\it Nu}{\it Ra}^{{\beta}_{\it Nu}}$ and ${\it Re} \approx {\alpha}_{\it Re}{\it Ra}^{{\beta}_{\it Re}}$. 
 
 For flows with ${\it Ha} = 0$, some disagreement with the data of \citet{Cioni00} is found (see figure \ref{fig3}a). At the same time, the computed value of ${\beta}_{\it Nu}$ is in a good agreement with the value $0.29 \pm 0.01$ for ${\Pran} = 0.025$ found in the experiments with the largest range $2 \times 10^5 \le {\it Ra} \le 8 \times 10^{10}$ by \citet{Glazier99} performed at $\Gamma = 0.5, 1, 2$. The exponent ${\beta}_{\it Re}$ is consistent with $0.46 \pm 0.02$ for ${\Pran} = 0.025$ and $0.42 \pm 0.03$ for ${\Pran} = 0.029$ found in measurements by \citet{Takeshita96} and \citet{Zurner19_1}, respectively, and with $0.45 \pm 0.04$ for ${\Pran} = 0.021$ in numerical simulations by \citet{Scheel17}. 
 
 An imposed magnetic field reduces the rate of heat transfer and kinetic energy in the entire studied range of {\it Ra}. At the same time, growth of ${\it Nu}$ and $\Rey$ becomes faster. In particular, the slope ${\beta}_{\it Nu}$ increases from $0.298$ at ${\it Ha} = 0$ to about $0.574$ at ${\it Ha} = 1400$. Similar increase is observed for ${\beta}_{\it Re}$ (see table \ref{table1} and figure \ref{fig3}). The results are consistent with the data of \citep{Cioni00}, the only experiment in the interesting for us range of ${\it Ra}$ and ${\it Ha}$.


\begin{table}
  \begin{center}
\def~{\hphantom{0}}
  \begin{tabular}{ccccccccccccc}
      ${\it Ra}$ & ${\it Ha}$ & ~   ${\alpha}_{\it Nu}$ & ${\beta}_{\it Nu}$ & ~ ${\alpha}_{\it Re}$ & ${\beta}_{\it Re}$ \\[3pt]
      
   
	$10^7 - 10^9$ & 0 & ~  $0.0819$ & $0.298 \pm 0.005$ & ~ $4.427$ & $0.457 \pm 0.006$ \\
	$10^7 - 10^9$ & 450 & ~  $0.0064 $ & $0.420  \pm  0.008$ & ~ $0.192 $ & $0.586  \pm  0.010$ \\
	$10^7 - 10^9$ & 650 & ~  $0.0019$ & $0.476  \pm  0.007$ & ~ $0.139$ & $0.591  \pm  0.018$ \\
	$10^7 - 10^9$ & 850 & ~  $ 0.0008$ & $0.513  \pm 0.009$ & ~ $0.076$ & $0.617 \pm 0.012$ \\
	$10^8 - 10^9$ & 1400 & ~  $0.0002$ & $0.574  \pm 0.009$ & ~ $0.054$ & $0.624 \pm 0.033$ \\

  \end{tabular}
  \caption{Scaling coefficients in the approximations ${\it Nu} \approx {\alpha}_{\it Nu}{\it Ra}^{{\beta}_{\it Nu}}$ and ${\it Re} \approx {\alpha}_{\it Re}{\it Ra}^{{\beta}_{\it Re}}$ based on the data of the present simulations (see figure \ref{fig3}).}
  \label{table1}
  \end{center}
\end{table}

\begin{figure}
	\centering 
	
\begin{tikzpicture}

\node (img4_1) {\includegraphics[scale=0.275]{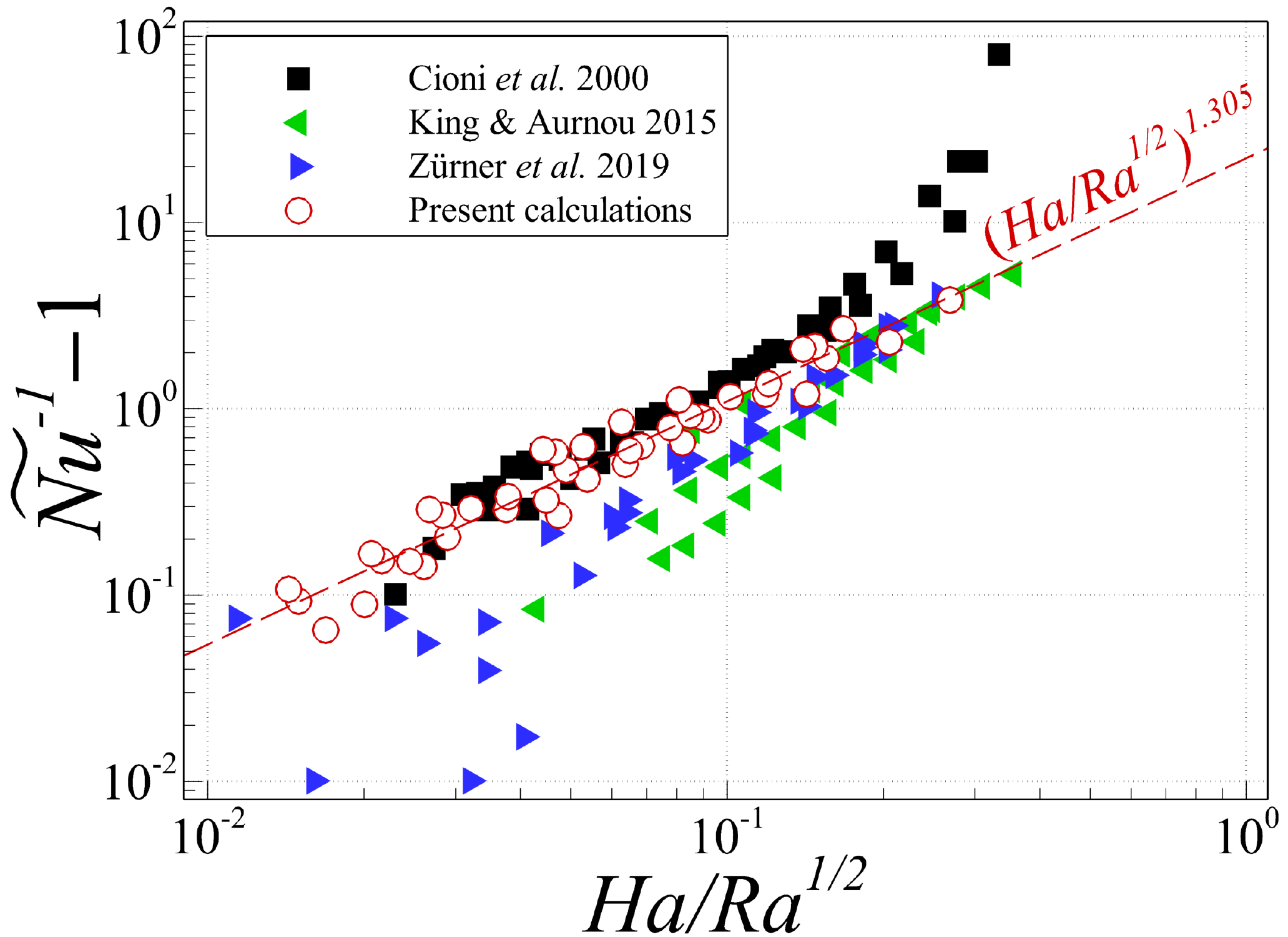}};
  	\node[left=of img4_1, xshift=6.5cm ,yshift=-1.0cm,rotate=0,font=\color{black}] {({\it a})};
\node [right=of img4_1, xshift=-0.75cm, yshift=0.00cm]  (img4_2)  {\includegraphics[scale=0.275]{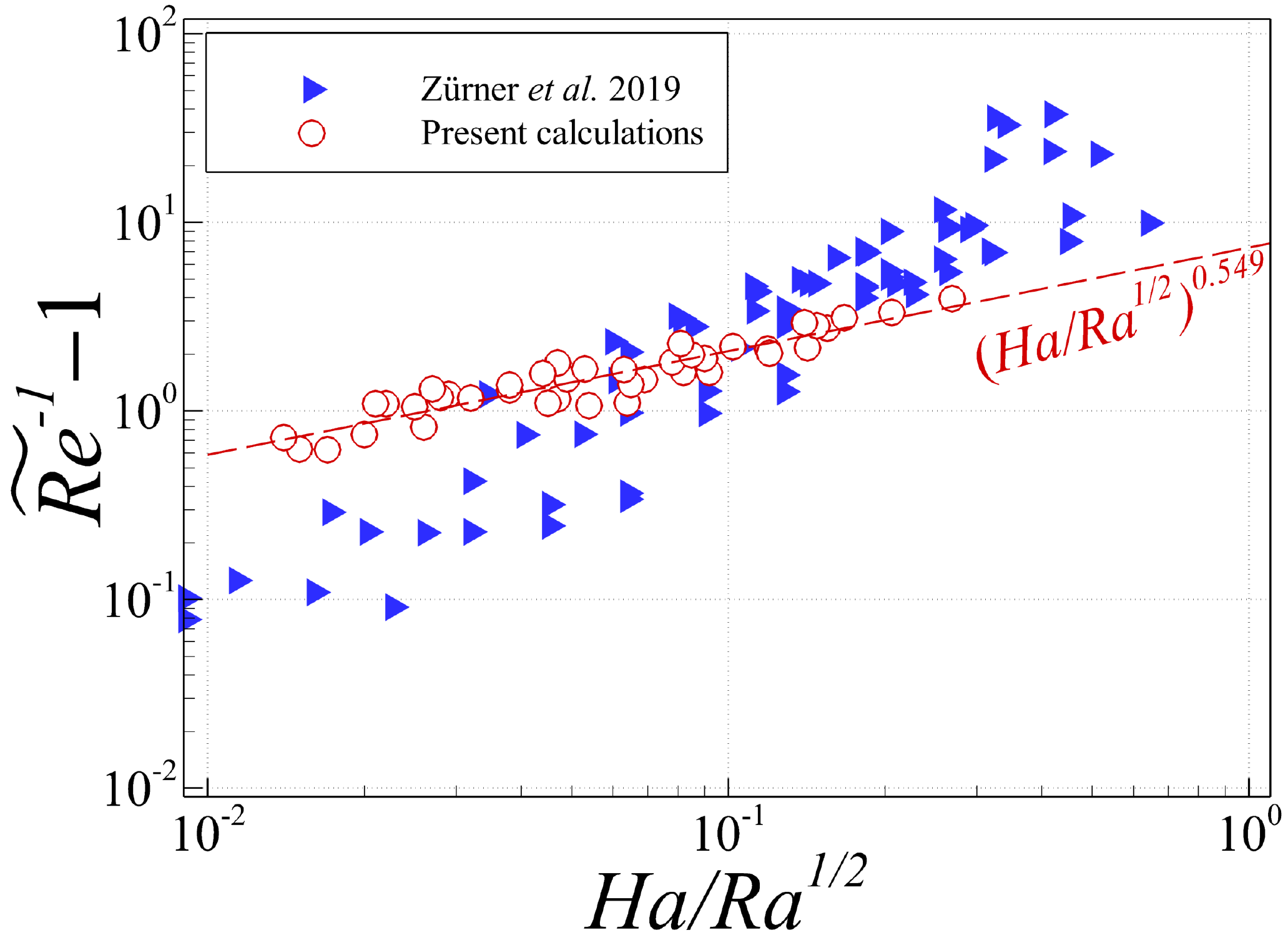}};
  	\node[left=of img4_2, xshift=6.5cm ,yshift=-1.0cm,rotate=0,font=\color{black}] {({\it b})};

\end{tikzpicture}

  \caption{ Normalized Nusselt number ${\it \widetilde{Nu}}$  vs ${\it Ha}/{\it Ra}^{1/2}$ ({\it a}) and Reynolds number ${\it \widetilde{Re}}$ vs  ${\it Ha}/{\it Ra}^{1/2}$  ({\it b}). Experimental data $\Gamma = 1$ are shown for comparison. Slope lines $ \backsim ({\it Ha}/{\it Ra}^{1/2})^{\beta}$ are plotted for guidance.}
  
\label{fig4}  
\end{figure}

The Nusselt  ${\it \widetilde{Nu}}=({\it Nu}-1)/({\it Nu}_{{\it Ha}=0}-1)$ and Reynolds ${\it \widetilde{Re}} = {\Rey}/{\Rey}_{{\it Ha}=0}$ numbers normalized by their reference values at ${\it Ha} = 0$  are shown in figure \ref{fig4}. The universal scaling behaviour of ${\it \widetilde{Nu}} = 1/(1 + {\alpha}_{\it \widetilde{Nu}}({\it Ha/Ha_c})^{\beta_{\it \widetilde{Nu}}})$ and ${\it \widetilde{Re}} = 1/(1 + {\alpha}_{\it \widetilde{Re}}({\it Ha/\sqrt{Ha_c}})^{\beta_{\it \widetilde{Re}}})$ was proposed by \citet{Zurner19_2}. To verify this hypothesis, we plot $1/{\it \widetilde{Nu}}-1$ and $1/{\it \widetilde{Re}}-1$ vs ${\it Ha/Ra^{1/2}}$ in logarithmic scales. The computed data provide $\beta_{\it \widetilde{Nu}} = 1.305 \pm 0.049$ and $\beta_{\it \widetilde{Re}} = 0.549 \pm 0.034$.

The qualitative agreement between the simulations and the experiments, especially with the experiments by \citet{Cioni00}, is observed for ${\it \widetilde{Nu}}$ at ${\it Ra} > {\it Ra}_c$ (${\it Ha}/{\it Ra}^{1/2} \sim 0.1$) and ${\it Ra} \gg {\it Ra}_c$ (${\it Ha}/{\it Ra}^{1/2} < 0.1$) in figure \ref{fig4}a. The results indicate that universal power is approached for flows at high Rayleigh numbers. At the same time, the standard error is higher and coefficient of determination ($R^2$) is lower than for the scaling ${\it Nu} \sim {\it Ra}^{{\beta}_{\it Nu}}$ presented above ($R^2 = 0.943$ in comparison to $R^2 = 0.998$ for ${\it Nu} \sim {\it Ra}^{\beta_{\it Nu}}$ at ${\it Ha} = 1400$). Similar situation is observed for the normalized Reynolds numbers in figure \ref{fig4}b. 

We expect an increase of the exponents $\beta_{\it \widetilde{Nu}}$ and $\beta_{\it \widetilde{Re}}$ in the wall mode regime at  ${\it Ra} \sim {\it Ra}_c$ (${\it Ha}/{\it Ra}^{1/2} \sim 0.25$), however the measurements of \citet{Zurner19_2} for $1/{\it \widetilde{Re}}-1$ show higher slope line $\sim ({\it Ha/Ha_c})^{1.73 \pm 0.05}$ for a wide range of ${\it Ra}$ and ${\it Ha}$. The discrepancy with this result and the presence of even stronger deviations in experimental data for moderate ${\it Ra}$ can be attributed to the limitations of the ultrasound Doppler velocimetry (UDV) used to probe the flow field discussed by \citet{Zurner19_2}. 

The estimates of the thickness of the thermal boundary layer are presented in the supplementary materials. The common slope method used in the experiments \citep{Takeshita96} and simulations \citep{Liu18} is applied, i.e. the intersection  point of the tangent of the time-averaged mean temperature profile near the sidewall and the horizontal line drawn through the mean value is taken. We only mention here that the results of the present simulations deviate by not more than $2 \%$ from the theoretical value $\delta_T \approx 1/(2\it Nu)$ of \citet{Grossmann01}, even in the presence of magnetic field.
 

 \section{\textbf{Concluding remarks}}
 
 We have performed direct numerical simulations of Rayleigh-Bénard convection in a cylinder with imposed vertical magnetic field. The range of high ${\it Ra}$ and ${\it Ha}$ never previously explored in numerical simulations was considered. The computations were performed on large grids with adequate resolution of internal features and boundary layers.
 
The new regime in the form of nearly Q2D upward and downward planar jets originating at the sidewalls and extending into the flow domain is identified at ${\it Ra}$ larger, but not much larger than ${\it Ra}_c$. The structures are reminiscent of Q2D extended vortex sheets found in MHD turbulence (see, e.g., \citet{Zikanov98}). Our results also show existence of rotating tongue-like wall modes at ${\it Ra} \rightarrow {\it Ra}_c$, in apparent qualitative similarity with rotating RBC \citep{Ecke92,Zhang19}.

The results of our DNS are consistent with available experimental and numerical data. In particular, we find faster growth of ${\it Nu}$ and ${\Rey}$ with ${\it Ra}$ in flows with strong magnetic fields. This effect can only be plausibly attributed to the formation of coherent large-scale Q2D structures in the flow field (see section \ref{results1}). The scaling laws for normalized Nusselt and Reynolds numbers reveal the same tendency as the experimental data, namely that the global transport properties approach a universal power law at ${\it Ra} \gg {\it Ra}_c$.

The presented work is considered by the authors as the beginning of a larger and more detailed study. Future investigations of the system are undoubtedly warranted. In particular, it would be interesting to further analyze the properties and physical mechanisms of the newly discovered Q2D regime, the behaviour of wall modes, the transport properties at ${\it Ra} > 10^9$ and ${\it Ha} > 1400$, and, finally, the apparent, albeit evidently incomplete, similarities between RBC with magnetic field and rotation.

 \section*{\textbf{Acknowledgments}}
 
Financial support is provided by the US NSF (Grant CBET 1803730) and the DFG grant KR 4445/2 $-$ 1. Computer time is provided by the Computing Center of the Technische Universit\"{a}t Ilmenau and the Leibniz Rechenzentrum Garching within Large Scale project pr62se.

\bibliographystyle{Akhmedagaev-bibst}
\bibliography{Akhmedagaev-bib}

\end{document}